\documentclass[reprint,amsmath,amssymb,aps,prx]{revtex4-2}
\usepackage{graphicx}
\usepackage{color}
\usepackage{amsmath}
\usepackage{amssymb}
\usepackage{amsthm}
\usepackage{extarrows}
\usepackage{booktabs}
\usepackage{float}
\usepackage[dvipdfm,colorlinks,urlcolor=blue,linkcolor=blue,anchorcolor=blue,citecolor=blue]{hyperref}
\usepackage{lineno}

\begin{document}

\title{From Near-Integrable to Far-from-Integrable: A Unified Picture of Thermalization and Heat Transport}

\author{Weicheng Fu$^{1,2,3}$}
%\email{fuweicheng@tsnu.edu.cn}
\thanks{These authors contributed equally to this work.}
\author{Zhen Wang$^{4}$}
\thanks{These authors contributed equally to this work.}
\author{Yisen Wang$^{3}$}
\author{Yong Zhang$^{5,3}$}
\email{yzhang75@xmu.edu.cn}
\author{Hong Zhao$^{5,3}$}
\email{zhaoh@xmu.edu.cn}

\affiliation{
$^1$ Department of Physics, Tianshui Normal University, Tianshui 741001, Gansu, China\\
$^2$ Key Laboratory of Atomic and Molecular Physics $\&$ Functional Material of Gansu Province, College of Physics and Electronic Engineering, Northwest Normal University, Lanzhou 730070, China\\
$^3$ Lanzhou Center for Theoretical Physics, Lanzhou University, Lanzhou 730000, Gansu, China\\
$^4$ Institute of Theoretical Physics, Chinese Academy of Sciences, Beijing 100190, China\\
$^5$ Department of Physics, Xiamen University, Xiamen 361005, Fujian, China
}

\date{\today }

\begin{abstract}

Whether and how a system approaches equilibrium is a central issue in nonequilibrium statistical physics, underpinning our understanding of both thermalization and heat transport. Bogoliubov's three-stage (initial,  kinetic, and hydrodynamic) evolution hypothesis provides a qualitative framework, but quantitative advances have mostly been restricted to near-integrable systems such as dilute gases.
In this work, we investigate the relaxation dynamics of a one-dimensional diatomic hard-point (DHP) gas and present a phase diagram that comprehensively characterizes relaxation behavior across the full parameter space, from near-integrable to far-from-integrable regimes. We first analyze thermalization (local energy relaxation in nonequilibrium states) and identify three universal dynamical regimes:
(i) In the near-integrable regime, kinetic processes dominate (initial + kinetic phase), local energy relaxation decays exponentially, and the thermalization time $\tau$ scales with perturbation strength $\delta$ as $\tau \propto \delta^{-2}$.
(ii) In the far-from-integrable regime, hydrodynamic effects dominate (initial + hydrodynamic phase), local energy relaxation follows power-law decay, and thermalization time scales linearly with system size $N$.
(iii) In the intermediate regime, Bogoliubov phase emerges, marked by the transition from kinetic to hydrodynamic relaxation.
The phase diagram further reveals that hydrodynamic behavior can emerge even in small systems when the system is sufficiently far from the integrable regime, challenging the conventional view that such effects are only significant in large systems. In the thermodynamic limit, the system's relaxation behavior depends on the order in which the limits ($N \to \infty$ or $\delta \to 0$) are taken.
We then turn to heat transport (decay of heat-current fluctuations in equilibrium) and demonstrate that its behavior is fully consistent with the thermalization process, leading to the first unified theoretical description of thermalization and heat transport.
Our approach provides a pathway for studying relaxation dynamics in a wide class of many-body systems, including quantum systems.

\end{abstract}

\maketitle

\section{Introduction}

Uncovering the microscopic origins of macroscopic relaxation remains a central objective of statistical physics. \cite{pathria2021statistical}. At the heart of this investigation lies the concept of \emph{thermalization}---the process by which isolated systems evolve to exhibit thermal equilibrium behavior in their measurable properties, and the study of the dynamical mechanisms that underlie this emergence of statistical typicality \cite{FORD1992271,10.1063/1.881363,RevModPhys.71.S346}.

The first numerical study of classical thermalization dates back to the 1950s, with the Fermi-Pasta-Ulam-Tsingou (FPUT) simulations revealing the now-famous FPUT recurrence, instead of the expected thermalized state \cite{Fermi1955,dauxois:ensl-00202296,2008LNP728G}. Over the ensuing decades, research has continued to reveal important insights into thermalization dynamics \cite{doi:10.1063/1.1861554,doi:10.1063/1.1889345,doi:10.1063/1.1855036,doi:10.1063/1.1849131,doi:10.1063/1.1858115,doi:10.1063/1.1861264,2008LNP728G}. Recent advancements have led to the discovery of a universal scaling law for large systems. In particular, near-integrable systems exhibit kinetically dominated thermalization, with the thermalization time $\tau$ scaling inversely with the square of the nonintegrability strength $\delta$, i.e.,  $\tau\propto\delta^{-2}$ \cite{PhysRevE.100.010101,Fu_2019,PhysRevE.100.052102,Onorato2019,PhysRevLett.124.186401, PhysRevE.104.L032104,Feng_2022,onorato2023wave,Wang_2024CTP}, a behavior that is independent of spatial dimensionality \cite{Wang24PRL}. Exceptions to this scaling have been observed in small or disordered systems, where thermalization may be threshold-dependent or exhibit distinct scaling behaviors \cite{PhysRevE.95.060202,Peng_2022,fu2025multitype,lin2024}.

In the quantum domain, thermalization has been explored in parallel \cite{annurev2015,Eisert2015,Gogolin_2016,Ueda2020,Rigol2008,science.1224953,PhysRevLett.125.240604,PhysRevLett.129.040602,PhysRevLett.129.063901,PhysRevX.12.031026,Le2023,PhysRevLett.132.027101,Andersen2025}. Landmark experiments such as the ``quantum Newton cradle'' \cite{Kinoshita2006} have demonstrated the suppression of thermalization in ultracold atomic gases. Theoretical frameworks, including the kinetic theory \cite{D'Alessio03052016,PhysRevLett.120.070603} and Fermi's golden rule \cite{PhysRevX.9.021027,PhysRevResearch.2.022034,PhysRevB.104.184302,PhysRevX.15.010501,PhysRevB.82.104306}, have revealed a same scaling law $\tau\propto\delta^{-2}$. However, there are exceptions where certain perturbations impede thermalization \cite{Mori_2018,PhysRevLett.106.040401,PhysRevLett.118.190601,PhysRevLett.127.130601}, such as integrability-breaking KAM-like perturbations \cite{PhysRevLett.127.130601}, and cases where anomalous scaling occurs \cite{Abanin2017,PhysRevResearch.5.043019}, underscoring the need for system-specific studies.
The striking similarities in thermalization dynamics between classical and quantum systems highlight their fundamental connections and suggest that advances in one domain may inform the other, offering valuable insights for the development of a unified theoretical framework of thermalization.

The universally observed inverse-square law of thermalization provides a clear understanding of thermalization mechanisms in near-integrable systems, both in classical \cite{PhysRevE.100.010101,Fu_2019,PhysRevE.100.052102,Onorato2019,PhysRevLett.124.186401, PhysRevE.104.L032104,Feng_2022,onorato2023wave,Wang_2024CTP,Wang24PRL} and quantum systems \cite{D'Alessio03052016,PhysRevX.9.021027,PhysRevResearch.2.022034,PhysRevB.104.184302,PhysRevX.15.010501,PhysRevLett.120.070603}.
However, thermalization dynamics in the far-from-integrable regimes remain underexplored. A unified framework for thermalization dynamics across the entire parameter space---from near-integrable to far-from-integrable regimes---remains a central challenge in nonequilibrium physics. This work provides critical advances toward solving this fundamental problem.

On the other hand, the fluctuation-dissipation theorem establishes a link between the relaxation processes of a system---particularly those in the linear response regime---and its transport properties \cite{R_Kubo_1966}. Significant progress in thermal transport research over the past fifty years has shown that heat conduction is determined by both model-specific properties (such as momentum conservation and the symmetry of the interparticle interaction potential (IIP)) and spatial dimensionality, due  to hydrodynamic effects \cite{lepri2003thermal,Dhar08AdvPhys,Lepri2016}. For example, hydrodynamic theory predicts that in one-dimensional (1D) momentum-conserving systems, heat conductivity $\kappa$ follows a scaling law with system size $N$ as $\kappa \propto N^\alpha$ ($0<\alpha<1$), resulting in anomalous heat conduction that violates the Fourier's law \cite{PhysRevLett.89.200601,PhysRevLett.108.180601,PhysRevLett.111.230601,Spohn2014}.
Specifically, for systems with asymmetric IIP, $\alpha = 1/3$ \cite{PhysRevLett.89.200601,PhysRevLett.108.180601,PhysRevLett.111.230601,Spohn2014,PhysRevE.73.060201,Delfini_2007,PhysRevE.89.022111}, while for symmetric IIP, it remains under debate whether $\alpha = 1/2$ or $\alpha = 2/5$
\cite{S.Lepri_1998,PhysRevE.58.7165,PhysRevLett.92.074302,PhysRevE.68.056124,PhysRevLett.125.024101,2024JPSJ.93.053001}. Further research has highlighted the role of kinetic effects in heat conduction in real systems with finite sizes \cite{PhysRevE.85.060102,Zhong_2013,Chen_2016,Jiang_2016}.
In asymmetric IIP models, when the nonintegrability strength is small, size-independent heat conduction behavior is observed due to prolonged kinetic processes, restoring the Fourier's law \cite{PhysRevE.90.032134,PhysRevE.97.010103,PhysRevLett.125.040604}. Despite these advancements, constructing a microscopic transport theory that spans the entire parameter space remains challenging.
Establishing quantitative connections between microscopic dynamics and macroscopic transport behavior, and exploring their links to thermalization dynamics, are additional core motivations of this study.

In this work, we take the one-dimensional diatomic hard-point (DHP) gas model \cite{casati1986energy} as an illustrating example to study thermalization dynamics and heat transport across regimes from near-integrable to strongly nonintegrable. Specifically, we investigate thermalization, through the relaxation of local energy in nonequilibrium states, and heat transport, via the decay of equilibrium heat current fluctuations. The DHP gas usually serves as an ideal testbed for exploring fundamental questions in nonequilibrium statistical physics \cite{casati1976computer,PhysRevE.89.042918, Boozer_2011,PhysRevE.84.031127, PhysRevLett.86.3554, casati1986energy,PhysRevLett.86.5486, PhysRevLett.89.180601, PhysRevE.67.015203, PhysRevLett.94.244301, PhysRevE.90.032134,PhysRevE.97.010103,PhysRevLett.125.040604,hurtado2020simulations, PhysRevE.90.012147, PhysRevLett.126.244503}.

Through analytical arguments and high-precision numerical simulations, we obtain a complete phase diagram that characterizes the relaxation dynamics of the DHP model as a function of its nonintegrability strength. This diagram reveals three distinct relaxation patterns for a given system size $N$:
(i) In the near-integrable regime, relaxation is dominated by kinetic processes. For thermalization, local energy decays exponentially with a characteristic timescale $\tau \propto \delta^{-2}$, independent of $N$. This behavior aligns with previous findings in classical lattice systems \cite{PhysRevE.100.010101,Fu_2019,PhysRevE.100.052102,Onorato2019,PhysRevLett.124.186401, PhysRevE.104.L032104,Feng_2022,onorato2023wave,Wang_2024CTP,Wang24PRL} and 1D Bose gases \cite{D'Alessio03052016,PhysRevX.9.021027,PhysRevResearch.2.022034,PhysRevB.104.184302,PhysRevX.15.010501,PhysRevLett.120.070603}. For heat conduction, the heat current autocorrelation function (HCAF) also decays exponentially with the same scaling, resulting in normal (i.e., $N$-independent) heat conduction.
(ii) In the far-from-integrable regime, hydrodynamic processes dominate. Local energy exhibits a power-law decay, with the thermalization time scaling proportionally with $N$. The HCAF shows a similar power-law decay, leading to anomalous (i.e., $N$-dependent) heat conduction.
(iii) In the intermediate regime, the system enters the Bogoliubov's three-stage evolution phase, characterized by a crossover from exponential to power-law decay in both local energy relaxation and the HCAF, as the relaxation timescale increases.

Moreover, the phase diagram clearly indicates that in the far-from-integrable regime, hydrodynamic effects emerge prominently even in small systems, without requiring the large-$N$ limit.
In addition, we find that in the thermodynamic limit, the order of taking the limits $N \to \infty$ (the thermodynamic limit) and $\delta \to 0$ (the integrable limit) leads to markedly different relaxation dynamics: taking $N \to \infty$ first results in kinetically dominated behavior, whereas taking $\delta \to 0$ first yields hydrodynamic relaxation.
These results establish a unified framework for understanding relaxation dynamics in the DHP model across the entire spectrum of nonintegrability strengths.
Our findings resolve longstanding ambiguities in the DHP model---such as the parameter-dependent divergence exponents of heat conductivity \cite{hurtado2020simulations}---and provide a quantitative connection between integrability, thermalization, and heat transport.
This framework is broadly applicable to low-dimensional systems and offers critical insights into analogous phenomena in quantum systems \cite{RevModPhys.93.025003}.

The paper is organized as follows. Section~\ref{sec:2} introduces the DHP model, its key parameters, and the necessary definitions. Section~\ref{sec:3} presents the theoretical analysis of thermalization dynamics, supported by numerical simulations. Section~\ref{sec:4} examines heat transport within the same framework, combining theoretical insights and numerical results. Finally, Section~\ref{sec:5} provides a summary and discussion of the main findings.

\section{The DHP model}\label{sec:2}

The 1D DHP system consists of $N$ particles, labeled sequentially, with alternating masses given by $m_1 = 1 - \delta/2$ and $m_2 = 1 + \delta/2$, while maintaining unit mass density. Here, $\delta \in [0,2)$ represents the mass difference, and the system becomes integrable when $\delta = 0$, making $\delta$ the sole intrinsic parameter---referred to as the \emph{nonintegrability strength}. For a fixed mean number density of unity, $N$ simultaneously determines the system length. In this setup, the sound speed $c_{\rm s} = \sqrt{3k_B T}$ is derived, where $k_B$ is the Boltzmann constant (set to unity throughout) and $T$ is the absolute temperature \cite{wong2002handbook}. The unit of time in this model is $T^{-1/2}$ \cite{timeTemp}, and we set $T = 1$ throughout the analysis. Periodic boundary conditions are applied in all calculations.

Particles exhibit free motion except for elastic collisions with nearest neighbors. When a collision occurs between the $i$th and $(i+1)$th particles, their velocities are updated according to the following expression
\begin{equation}\label{eq_evolution_velocity}
\begin{bmatrix}
  \tilde{v}_i \\
  \tilde{v}_{i+1}
\end{bmatrix} = \frac{1}{2}
\begin{bmatrix}
m_i - m_{i+1} & 2m_{i+1} \\
2m_i & m_{i+1} - m_i
\end{bmatrix}
\begin{bmatrix}
v_i \\
v_{i+1}
\end{bmatrix},
\end{equation}
where the tilde denotes the post-collision velocities.

For the system with mean particle spacing unity, adjacent particles collide \emph{if and only if} their relative velocity $\nu = v_i - v_{i+1} > 0$. The mean free time between collisions, $\tau_{\rm f}$, is given by
\begin{equation}\label{eq-meanTime}
\tau_{\rm f} = \frac{1}{\int_{0}^{\infty}\nu f(\nu)d\nu} = \sqrt{\frac{\pi}{T}\left(1 - \frac{\delta^2}{4}\right)},
\end{equation}
where $f(\nu) = \frac{1}{\sqrt{2 \pi T / \mu}} e^{-\frac{\mu \nu^2}{2T}}$ is the distribution function, and $\mu = \frac{m_i m_{i+1}}{m_i + m_{i+1}} = \frac{1}{2} \left(1 - \frac{\delta^2}{4}\right)$ is the reduced mass of the colliding pair.

\begin{figure}[t]
  \centering
  \includegraphics[width=1\columnwidth]{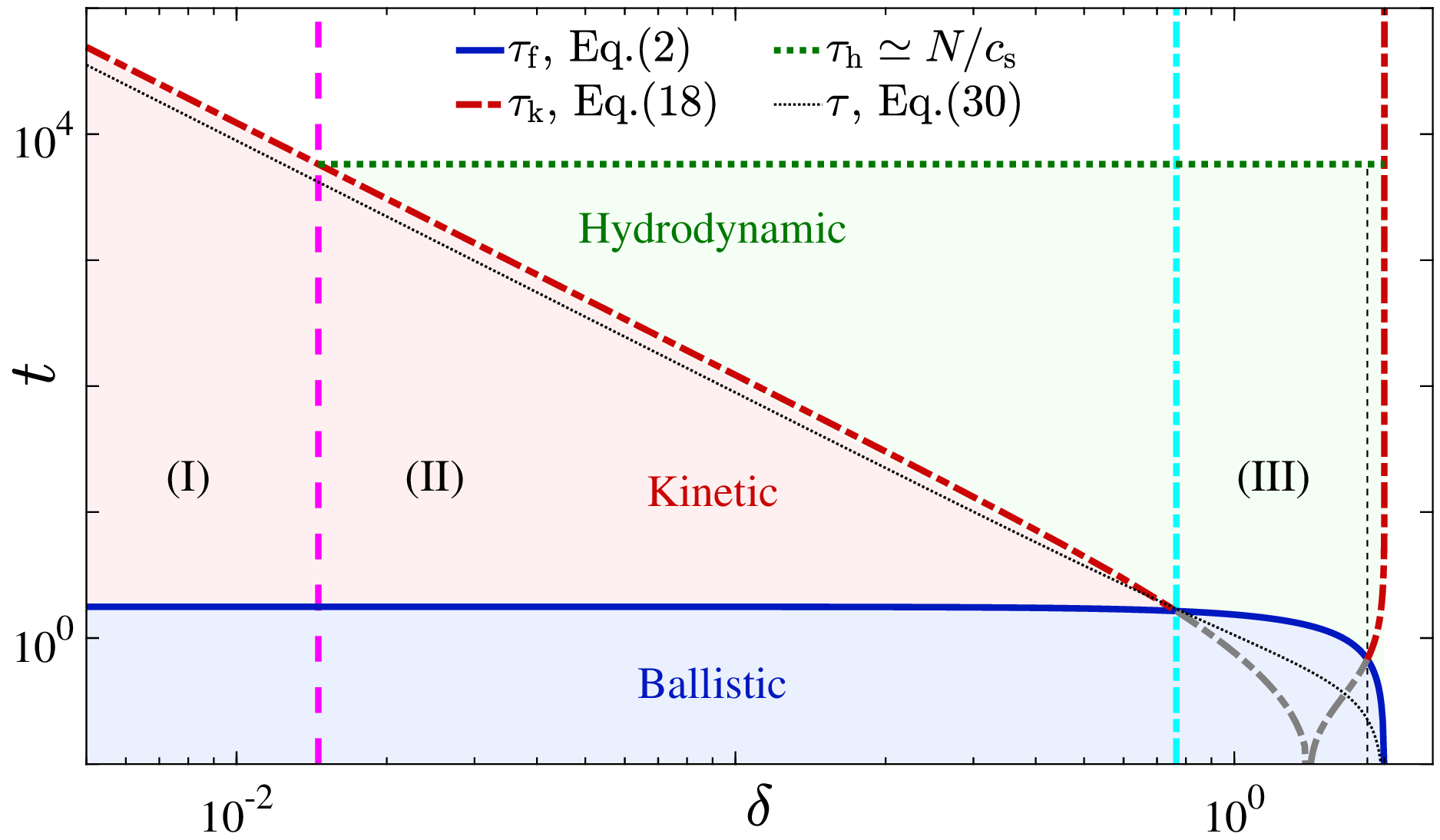}
  \caption{Diagram of characteristic timescales versus $\delta$ in a log-log scale, with $N = 10^4$ for demonstration. The curves corresponding to Eqs. (\ref{eq-Teq-delta}) and (\ref{eq-tau}) are shown for reference. An appropriate coefficient is introduced so that the three curves, $\tau_{\rm f}$, $\tau_{\rm k}$, and $\tau$, intersect at the point $\delta = \sqrt{2 - \sqrt{2}}$, corresponding to $|\lambda|=g$.  The left magenta vertical dashed line marks the intersection of $\tau_{\rm f}$ and $\tau_{\rm h}$, while the right cyan vertical line marks the intersection of $\tau_{\rm f}$ and $\tau_{\rm k}$. For specific details, see the discussion in the main text.}\label{fig-taus}
\end{figure}

From Eq. (\ref{eq-meanTime}), we observe that $\tau_{\rm f}$ approaches $\sqrt{\pi/T}$ (a constant) as $\delta \to 0$, while $\tau_{\rm f} \to 0$ as $\delta \to 2$ (i.e., when $m_1 \to 0$ and $m_2 \to 2$), as shown by the blue curve in Fig.~\ref{fig-taus}. In equilibrium, the equipartition theorem gives $\langle m_1 v_1^2 \rangle = \langle m_2 v_2^2 \rangle = k_B T$, which leads to $\langle v_1^2 \rangle \to \infty$ and $\langle v_2^2 \rangle \to T/2$ as $\delta \to 2$. This implies that light particles become ultra-fast and rebound with divergent velocities, causing $\tau_{\rm f} \to 0$. Although physical speeds cannot exceed the speed of light, the velocities near $\delta \to 2$ would require relativistic corrections. This extreme regime introduces significant complexity in the analysis of relaxation dynamics and transport properties, as relativistic effects become non-negligible. A more detailed exploration of these relativistic aspects will be addressed in future work \cite{Fudmass2}.

Inspired by the Bogoliubov hypothesis \cite{bogoliubov1962problems,PhysRevA.31.1883,mitropolskii1993nn,Bogolyubov_Jr_1994}, we consider the energy relaxation process in gases, which exhibits hierarchical behaviors---ballistic, kinetic, and hydrodynamic, as illustrated in Fig.~\ref{fig-taus}, specifically in region (II). For times $0 < t < \tau_{\rm f}$, particles primarily exhibit \emph{inertial motion}, leading to \emph{ballistic behavior}. After $\tau_{\rm f}$, \emph{local two-body collisions} result in \emph{kinetic} relaxation. As the time progresses toward the macroscopic relaxation time $\tau_{\rm h} \sim N / c_{\rm s}$, \emph{hydrodynamic} effects, characterized by the \emph{collective motion} of particles, dominate the relaxation behavior. Therefore, $\tau_{\rm f}$ marks the onset of the kinetic regime, while the characteristic time $\tau_{\rm k}$ for the relaxation of a specific physical quantity (such as the fluctuation of local energy or heat current), which defines the end of the kinetic regime, requires further analysis.

To provide a unified framework for describing thermalization and heat transport, we categorize the following three cases:

$\bullet$ Case (i): $\tau_{\mathrm{k}} \gg \tau_{\mathrm{f}}$ and $\tau_{\mathrm{k}} \to \tau_{\mathrm{h}}$ (or even $\tau_{\mathrm{k}} > \tau_{\mathrm{h}}$), corresponding to region (I) to the left of the magenta vertical dashed line in Fig.~\ref{fig-taus}. Here, kinetic effects dominate, while hydrodynamic behavior is suppressed.

$\bullet$ Case (ii): $\tau_{\mathrm{f}} \ll \tau_{\mathrm{k}} \ll \tau_{\mathrm{h}}$, corresponding to region (II) between the magenta and cyan vertical lines in Fig.~\ref{fig-taus}. This intermediate regime is referred to as the Bogoliubov phase, where ballistic, kinetic, and hydrodynamic effects coexist.

$\bullet$ Case (iii): $\tau_{\mathrm{h}} \gg \tau_{\mathrm{k}}$ and $\tau_{\mathrm{k}} \to \tau_{\mathrm{f}}$, corresponding to region (III) to the right of the cyan vertical dashed line in Fig.~\ref{fig-taus}. In this regime, the kinetic contribution becomes negligible, and the system exhibits pronounced hydrodynamic behavior.

These three regimes will be referenced repeatedly in subsequent discussions. Physically, the characteristic time $\tau_{\mathrm{k}}$ is governed by the system's intrinsic nonintegrability and arises from \emph{local collisions}, with negligible dependence on system size $N$. In contrast, $\tau_{\mathrm{h}} \sim N / c_{\rm s}$ scales linearly with $N$, making it strongly size-dependent. We now proceed to investigate the behavior of $\tau_{\mathrm{k}}$ during thermalization and transport dynamics, thus elucidating their intrinsic connection.

\section{Thermalization}\label{sec:3}

In this section, we investigate the rate at which the system reaches thermal equilibrium and the underlying mechanisms. From Eq.~(\ref{eq_evolution_velocity}), it can be easily derived that the energy of colliding particles $i$ and $(i+1)$ evolves as
\begin{equation}\label{eq_evo_energy}
\begin{bmatrix}
  \tilde{\mathcal{E}}_i \\
  \tilde{\mathcal{E}}_{i+1}
\end{bmatrix} = \boldsymbol{A}
\begin{bmatrix}
  \mathcal{E}_i \\
  \mathcal{E}_{i+1}
\end{bmatrix} + G
\begin{bmatrix}
  1 \\
  -1
\end{bmatrix},
\end{equation}
where $\mathcal{E}_i = \frac{1}{2}m_i v_i^2$ is the kinetic energy of particle $i$, and
\begin{equation}\label{eq_transition_matrix}
\boldsymbol{A} = \frac{1}{4}
\begin{bmatrix}
  \delta^2 & 4 - \delta^2 \\
  4 - \delta^2 & \delta^2
\end{bmatrix} = \frac{1}{2}
\begin{bmatrix}
  1 & 1 \\
  1 & 1
\end{bmatrix}
+ \frac{\lambda}{2}
\begin{bmatrix}
  1 & -1 \\
  -1 & 1
\end{bmatrix}
\end{equation}
is the transition matrix, with eigenvalues
\begin{equation}\label{eq-lambda}
  1 \quad \text{and} \quad \lambda = \frac{\delta^2}{2} - 1 \in [-1, 1),
\end{equation}
and
\begin{equation}\label{eq_geometric_mean}
G = \frac{1}{2} (m_i - m_{i+1}) m_i m_{i+1} v_i v_{i+1} = \pm g \sqrt{\mathcal{E}_i \mathcal{E}_{i+1}}
\end{equation}
describes the \emph{correlation of velocities} (or the nonlinear coupling of energies) before the collision. Here,
\begin{equation}\label{eq-g}
g = \delta \sqrt{1 - \frac{\delta^2}{4}} \in [0, 1]
\end{equation}
characterizes the strength of velocity correlation between colliding particles. The sign ``$\pm$'' depends on ${\rm sgn}(m_i - m_{i+1}) {\rm sgn}(v_i v_{i+1})$, which indicates the type of collision. For example, a positive mass difference ($m_i - m_{i+1} > 0$) corresponds to a left-heavy/right-light particle pair, while a negative mass difference ($m_i - m_{i+1} < 0$) corresponds to a left-light/right-heavy configuration. Additionally, $v_i v_{i+1} > 0$ (or $< 0$) indicates a collision between particles moving in the same (or opposite) direction. In equilibrium, we have
\begin{equation}\label{eq-G-G2}
\langle G \rangle = 0 \quad \text{and} \quad \langle G^2 \rangle = \frac{g^2 T^2}{4}.
\end{equation}

\begin{figure}[t]
  \centering
  \includegraphics[width=1\columnwidth]{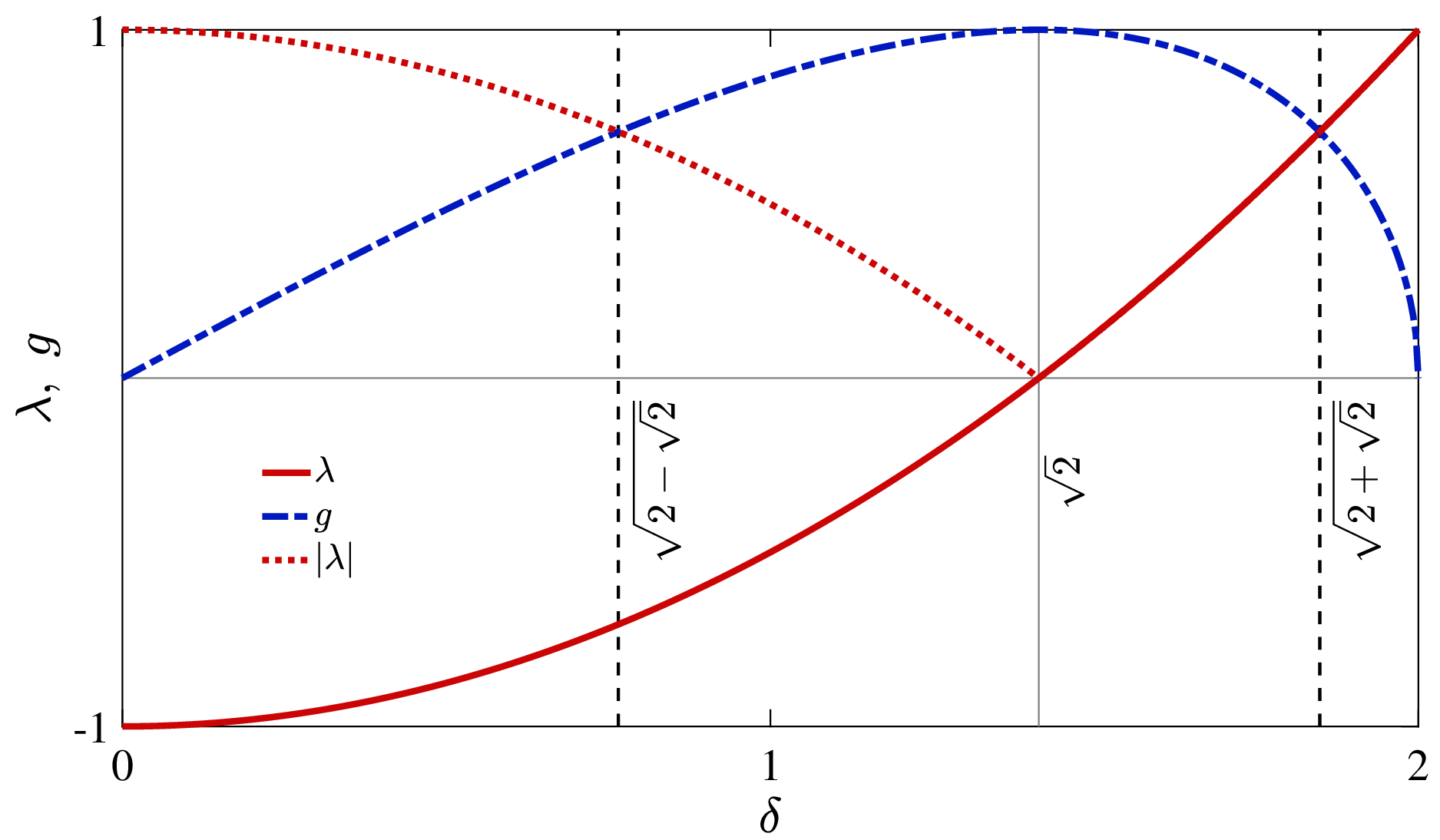}
  \caption{Dependence of $\lambda$ and $g$ on $\delta$. The vertical dashed lines correspond to $\delta_{\rm c}=\sqrt{2\mp\sqrt{2}}$, where $|\lambda|=g$.}\label{fig-lambda-g}
\end{figure}

From Eq.~(\ref{eq_evo_energy}), it is evident that the collision redistributes the energy of the particles through a combination of weighted algebraic mean (the first term on the right-hand side, linear) and geometric mean (the second term, nonlinear). To illustrate the role of these two terms, we now consider two special cases. First, in the case of $\delta = 0$, i.e., the integrable case, $\lambda = -1$ and $g = 0$, so $G = 0$. In this case, Eq.~(\ref{eq_evo_energy}) simplifies to
\begin{equation}\label{eq_evo_energy_d0}
\tilde{\mathcal{E}}_i = \mathcal{E}_{i+1}, \quad \text{and} \quad \tilde{\mathcal{E}}_{i+1} = \mathcal{E}_i,
\end{equation}
which shows that collisions merely exchange energy between particles without redistributing it. Thus, the integrable system does not spontaneously reach a thermalized state. Second, in the case of $\delta = \sqrt{2}$, we have $\lambda = 0$ and $g = 1$, so $G = \pm \sqrt{\mathcal{E}_i \mathcal{E}_{i+1}}$, and Eq.~(\ref{eq_evo_energy}) becomes
\begin{equation}\label{eq_evo_energy_dsq2}
\tilde{\mathcal{E}}_i = \bar{\mathcal{E}} + G, \quad \text{and} \quad \tilde{\mathcal{E}}_{i+1} = \bar{\mathcal{E}} - G,
\end{equation}
where $\bar{\mathcal{E}} = \frac{1}{2} \left( \mathcal{E}_i + \mathcal{E}_{i+1} \right)$ is the mean energy. The system can reach an equipartition state in a single collision when only the algebraic mean is considered. However, the geometric mean ($G$) prevents the system from staying in this state. Clearly, for $\delta \in (0, 2)$, we have $0 < |\lambda| < 1$ and $0 < |G| \leq \sqrt{\mathcal{E}_i \mathcal{E}_{i+1}}$, thus the weighted algebraic mean drives the system toward local equilibrium, while the geometric mean drives the system away from local equilibrium. Intuitively, $|\lambda|$ and $g$ roughly characterize the strength of these two competing effects, as illustrated in Fig.~\ref{fig-lambda-g}.
Let $|\lambda| = g$, we obtain
\begin{equation}\label{eq-delta-c}
\delta_{\rm c} = \sqrt{2 \mp \sqrt{2}}.
\end{equation}

\begin{figure*}[t]
  \centering
  \includegraphics[width=2\columnwidth]{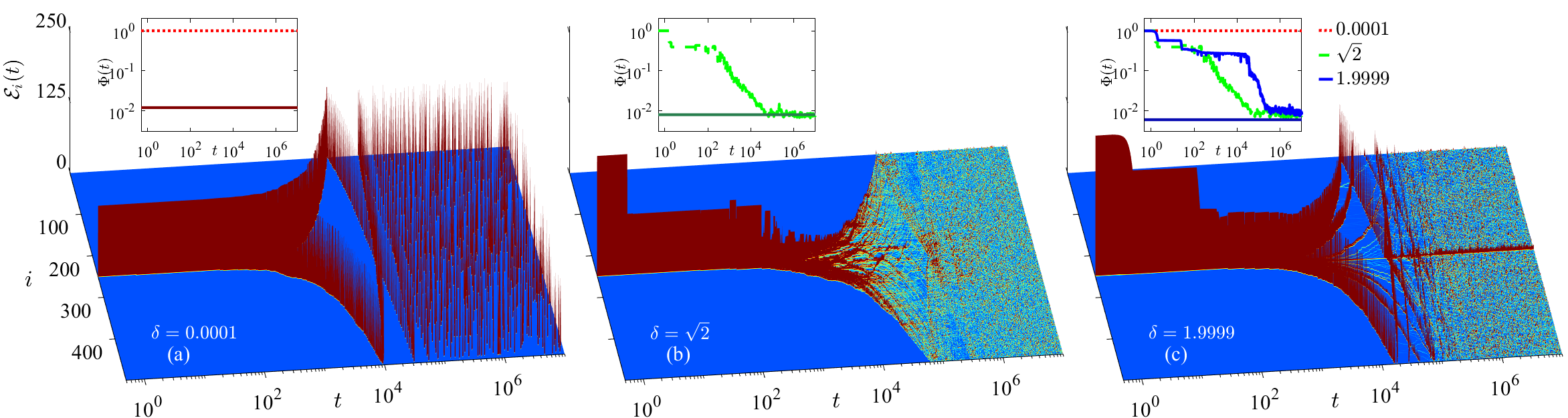}\\
  \caption{ Evolution of energy $\mathcal{E}_i(t)$ at the fixed $N=500$. Panels (a)-(c) correspond to $\delta=10^{-4}$, $\sqrt{2}$, and $1.9999$, respectively. Insets: Evolution of the IPR obtained from the data in the main panel. The horizontal lines in insets represent the theoretical values for the thermalized state, see Eq.~(\ref{eq_Phi_theroy}).
   }\label{fig-var}
\end{figure*}

\subsection{Estimation of timescale for thermalization }

In the kinetic theory of gases \cite{boltzmann1964lectures, chapman1990mathematical, ehrenfest1990conceptual, Brown2009}, the molecular chaos hypothesis, also known as the \emph{Stosszahlansatz}, asserts that the velocities of colliding particles are uncorrelated and independent of position, i.e., $\langle G \rangle = 0$. This implies that the collisions in the system are statistically independent, making each snapshot of collision statistically identical. Consequently, the system's evolution can be described by the dynamics of a pair of colliding particles. This leads to an evolution equation for the mean energy
\begin{equation}\label{eq-en}
\begin{bmatrix}
  \tilde{E}_1 \\
  \tilde{E}_2
\end{bmatrix} = \boldsymbol{A}
\begin{bmatrix}
  E_1 \\
  E_2
\end{bmatrix} \Rightarrow
\boldsymbol{E}(1) = \boldsymbol{A} \boldsymbol{E}(0),
\end{equation}
where $E_1 = \langle \mathcal{E}_i \rangle$ and $E_2 = \langle \mathcal{E}_{i+1} \rangle$ represent the ensemble averages over the initial state. Thus, the energy at time $t$ is given by
\begin{equation}\label{eq-matix-en-An}
\boldsymbol{E}(t) = \boldsymbol{A}^{t/\tau_{\rm f}} \boldsymbol{E}(0),
\end{equation}
where
\begin{equation}\label{eq-matix-en-AnCoef}
\boldsymbol{A}^{t/\tau_{\rm f}} = \frac{1}{2} \begin{bmatrix}
  1 &  1 \\
  1 &  1
\end{bmatrix}
+ \frac{\lambda^{t/\tau_{\rm f}}}{2} \begin{bmatrix}
  1 &  -1 \\
  -1 &  1
\end{bmatrix}.
\end{equation}
Since $|\lambda| < 1$ when $\delta \neq 0$, we have
\begin{equation}\label{eq-matix-en-AnCoef2}
\lim_{t \to \infty} \lambda^{t/\tau_{\rm f}} \to 0 \Rightarrow \lim_{t \to \infty} \boldsymbol{A}^{t/\tau_{\rm f}} = \frac{1}{2} \begin{bmatrix}
  1 &  1 \\
  1 &  1
\end{bmatrix},
\end{equation}
which results in
\begin{equation}\label{eq_E1_E2}
E_1(t) = E_2(t) = \frac{1}{2} \left[E_1(0) + E_2(0)\right],
\end{equation}
indicating that the system reaches an equipartition state as $t \to \infty$. The rate at which $\lambda^{t/\tau_{\rm f}}$ approaches zero reflects the speed at which the system enters this state. To characterize the relaxation behavior more clearly, we define
\begin{equation}\label{eq_E_Delta}
|\lambda|^{t/\tau_{\rm f}} = \exp(-t/\tau_{\rm k}),
\end{equation}
where $\tau_{\rm k}$ is the characteristic time of relaxation. From Eq.~(\ref{eq_E_Delta}), it follows that
\begin{equation}\label{eq-Teq-delta}
\tau_{\rm k} = -\frac{\tau_{\rm f}}{\ln(|\lambda|)} \propto
\begin{cases}
  \delta^{-2}, & \delta \to 0; \\
  (2 - \delta)^{-1/2}, & \delta \to 2,
\end{cases}
\end{equation}
providing an estimate for the timescale of thermalization, particularly during the stage dominated by local collisions. Equations~(\ref{eq_E_Delta}) and (\ref{eq-Teq-delta}) represent the main theoretical results of this work, demonstrating that interparticle collisions drive the gas toward local equilibrium in an exponential manner.

From Eq.~(\ref{eq-Teq-delta}), we observe that $\tau_{\rm k}$ reaches its minimum at $\delta = \sqrt{2}$, where $\lambda = 0$, meaning the system achieves equipartition after a single collision. However, as shown in Fig.~\ref{fig-lambda-g}, $g = 1$ peaks at $\delta = \sqrt{2}$, suggesting that the velocity correlation is strongest, and the effect of $G$, which drives the system away from local equilibrium, becomes significant, as seen in Eq.~(\ref{eq_evo_energy_dsq2}).

Physically, the analysis presented above applies primarily when $g \ll |\lambda|$. When $g > |\lambda|$, i.e., $\sqrt{2 - \sqrt{2}} < \delta < \sqrt{2 + \sqrt{2}}$, collective motion driven by spatial correlations becomes significant, leading to pronounced hydrodynamic behavior characterized by a long-time tail. Moreover, it is evident that the diagonal elements of the transition matrix $\boldsymbol{A}$ are smaller than the off-diagonal elements, i.e., $\delta^2 < (4 - \delta^2)$ for $\delta \in [0, \sqrt{2})$. This implies that after a collision, particles transfer more energy to each other than they retain, resulting in a \emph{weak memory effect}. Conversely, for $\delta \in (\sqrt{2}, 2)$, we have $\delta^2 > (4 - \delta^2)$, meaning that most of the energy stays with the colliding particles, resulting in a \emph{stronger memory effect}, which leads to more pronounced hydrodynamic behavior. These predictions will be validated through numerical simulations.

\subsection{Numerical verification of thermalization}

In this subsection, we perform molecular dynamics simulations to validate the theoretical analysis, particularly the estimation of the thermalization rate, as provided by Eq.~(\ref{eq-Teq-delta}). To this end, we utilize the inverse participation ratio (IPR), $\Phi(t)$, defined as

\begin{equation}
\Phi(t) = \frac{\sum_{i=1}^{N} \mathcal{E}_i(t)^2}{\sum_{i=1}^{N} \mathcal{E}_i(0)^2},
\end{equation}
which serves as a measure \cite{Wegner1980} of the evolution of local energy fluctuations over time \cite{IPR_Var}. For $\delta = 0$, the system is integrable, and the IPR remains constant, signifying no thermalization, consistent with ballistic motion. However, for $\delta \neq 0$, the IPR changes, indicating the onset of thermalization. The ensemble average $\langle \Phi(t) \rangle$ is expected to stabilize at a minimum value in the thermalized state. The system is evolved using the event-driven algorithm \cite{PhysRevE.67.015203}.

\begin{figure*}[t]
  \centering
  \includegraphics[width=2\columnwidth]{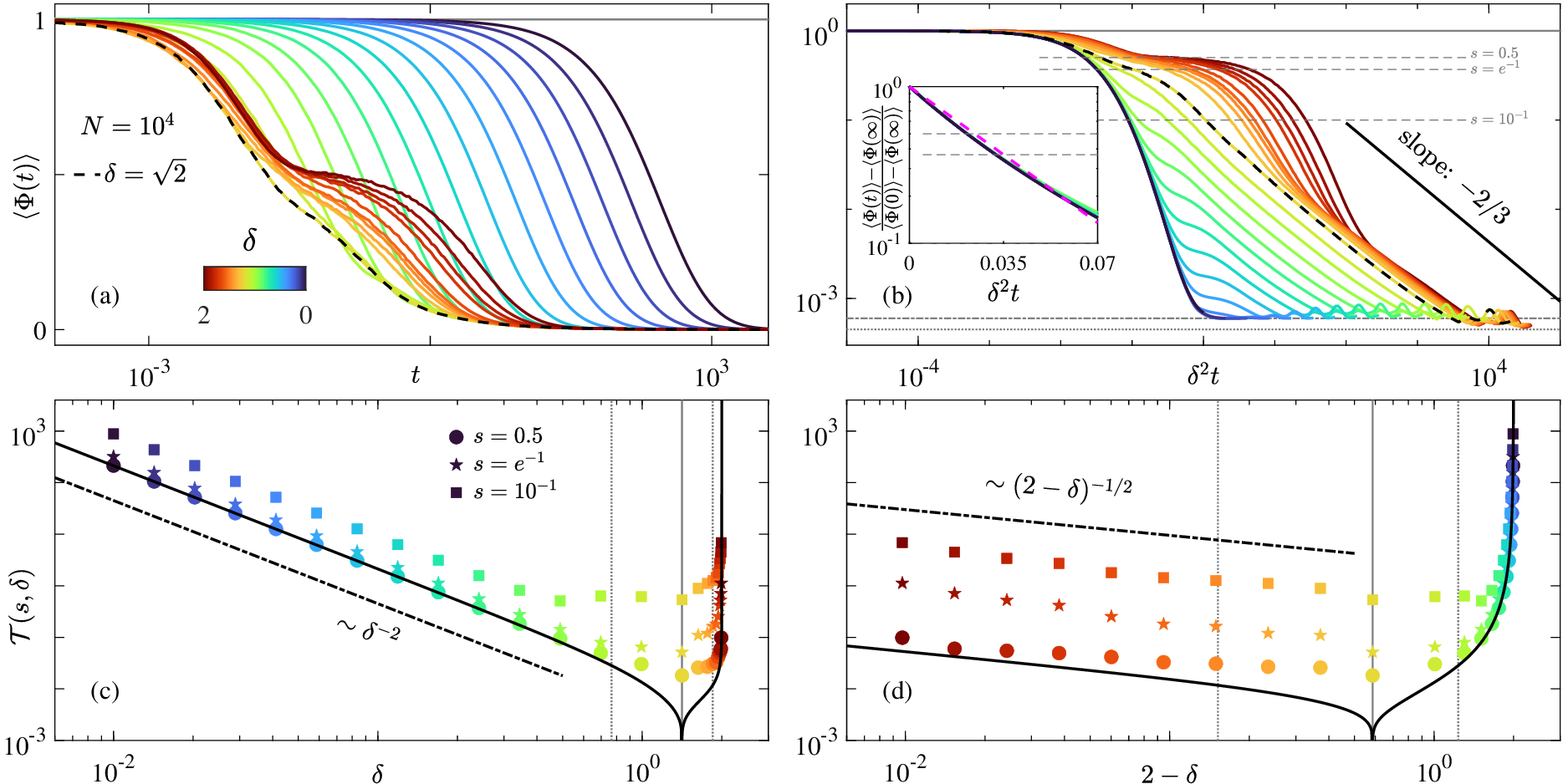}
  \caption{(a) IPR $\langle\Phi(t)\rangle$ for various $\delta$. (b) Same as panel (a), but with the horizontal axis rescaled by $\delta^2$ in a log-log scale. The two horizontal reference lines at the bottom correspond to the values of $6/N$ and $9/(2N)$ respectively. Inset: The ordinates are rescaled and plotted in a semi-logarithmic scale, with the magenta line representing the exponential decay function for reference. (c) $\mathcal{T}(s,\delta)$ as a function of $\delta$ for a given $s$, see the horizontal gray reference lines in panel (b). (d) Same as the panel (c), but the horizontal axis replotted as $2-\delta$.}\label{fig-Teq}
\end{figure*}

In the numerical simulations, the particle spacing is initialized randomly, and then two particles, $i$ and $j$, are randomly selected, setting the total energy of the system while ensuring momentum conservation, i.e.,
\begin{equation}\label{eq-initial-NT}
m_i v_i^2 + m_j v_j^2 = NT \quad \& \quad m_i v_i + m_j v_j = 0,
\end{equation}
while the remaining particles are kept at rest. Under these conditions, it is straightforward to derive that
\begin{equation}\label{eq_Phi_theroy}
  \langle \Phi(\mathcal{T}_{\rm eq}) \rangle = \frac{3(8 + \delta^2)}{N(4 + \delta^2)} \in \left( \frac{9}{2N}, \frac{6}{N} \right)
\end{equation}
in the thermalized state, where $\mathcal{T}_{\rm eq}$ represents the thermalization time, i.e., when $\langle \Phi(t) \rangle$ stabilizes to the corresponding value of Eq.~(\ref{eq_Phi_theroy}), signaling that the system has reached thermal equilibrium.

To visualize the system's dynamics, Fig.~\ref{fig-var}(a) shows the evolution of the particle energy for $\delta = 10^{-4}$ and $N = 500$, where the initial total energy is assigned to the 249th and 250th particles. The energy propagates almost ballistically, and the corresponding $\Phi(t)$ remains nearly constant within the computed time range (red dotted line in the inset). Specifically, $\Phi(t) \approx 1 \gg \langle\Phi(\mathcal{T}_{\rm eq})\rangle = 0.012$, indicating that the system  is still far from equilibrium, as it remains in the near-integrable regime, making thermalization difficult.

Figure~\ref{fig-var}(b) presents the results for $\delta = \sqrt{2}$, where energy rapidly transfers to adjacent particles, and $\Phi(t)$ decreases quickly (green dotted line in the inset), signifying the system's approach to equilibrium. Around $t \sim 10^6$, $\Phi(t)$ stabilizes at $\langle\Phi(\mathcal{T}_{\rm eq})\rangle = 0.008$ (horizontal line in the inset), suggesting that the system has reached thermal equilibrium.

Figure~\ref{fig-var}(c) shows the results for $\delta = 1.9999$. After a brief period (approximately $t \sim 10^4$), several well-defined energy wave packets form and propagate outward. After multiple collisions, these wave packets gradually dissipate, but a localized wave packet remains at the center. The system continues to evolve toward equilibrium, with $\Phi(t)$ slowly approaching the theoretical value $\langle\Phi(\mathcal{T}_{\rm eq})\rangle = 0.006$, as indicated by the blue line in the inset, with the other two lines shown for comparison.

To gain further insight into the system's thermalization, Fig.~\ref{fig-Teq}(a) shows $\langle \Phi(t) \rangle$ for various values of $\delta$ at the fixed $N = 10^4$, averaged over $10^4$ trajectories with random initialization to minimize fluctuations. The rate at which $\langle \Phi(t) \rangle$ approaches a stable value as $\delta$ increases is nonmonotonic, in qualitative agreement with the results in Fig.~\ref{fig-var}. The curve for $\delta=\sqrt{2}$ decays the fastest, as shown by the black dashed line.

Figure~\ref{fig-Teq}(b) reveals a collapse for small $\delta$ when $t$ is rescaled by $\delta^2$. The curve approximates exponential decay (see the inset in semi-logarithmic scale) toward its minimum value $6/N$, indicating that $\mathcal{T}_{\rm eq} \propto \tau_{\rm k} \propto \delta^{-2} \gg \tau_{\rm h}$ as $\delta \to 0$. This suggests that the system remains in region (I) (see Fig.~\ref{fig-taus}), where kinetics dominate.

As $\delta$ increases, a crossover occurs in $\langle \Phi(t) \rangle$ from exponential decay (typical of the kinetic process) to power-law decay (characteristic of the hydrodynamic process), signaling a transition from region (I) to region (II). Specifically, for larger $\delta$ (e.g., $\delta \to \sqrt{2}$), $\langle \Phi(t) \rangle \sim t^{-2/3}$ before stabilizing, indicating that the system enters region (III).

\begin{figure*}[t]
  \centering
  \includegraphics[width=2\columnwidth]{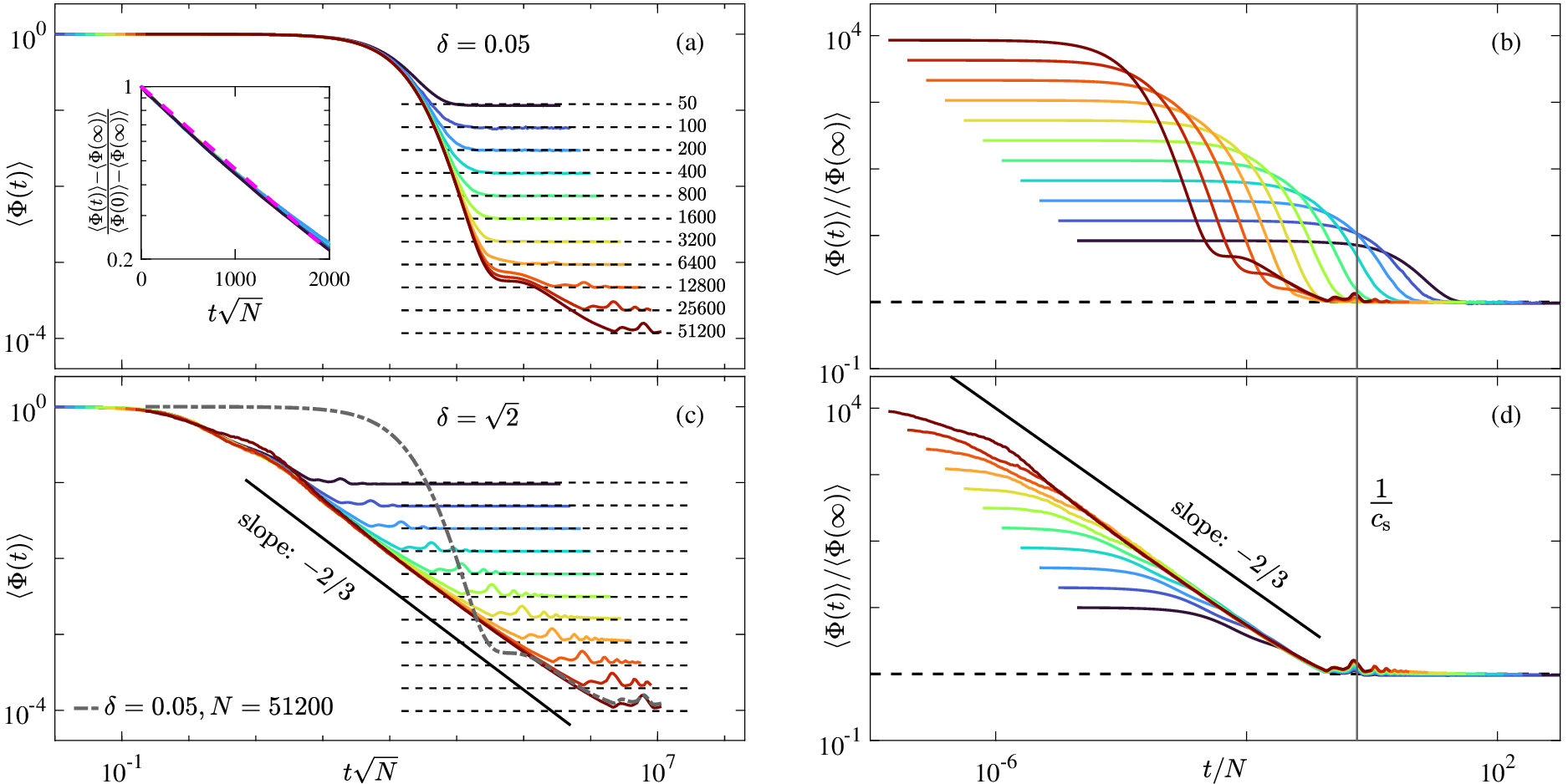}
  \caption{IPR $\langle\Phi(t)\rangle$ for various $N$ in a log-log scale. (a)-(b) for $\delta=0.05$, and (c)-(d) for $\delta=\sqrt{2}$. The horizontal panels differ only in the rescaling of the axes. All panels share the legend. Lines with different slopes are drawn for reference. }\label{fig-phi}
\end{figure*}

For a large $\delta$, particularly as $\delta \to 2$, $\langle \Phi(t) \rangle$ exhibits complex decay behavior: it rapidly decays to a small plateau, then decays again, and finally follows a power-law decay of $\langle \Phi(t) \rangle \sim t^{-2/3}$, approaching its stable value. This behavior suggests that for very large $\delta$, the thermalization process is dominated by hydrodynamic effects, as $\delta^2 \gg (4 - \delta^2)$ when $\delta \to 2$, reflecting a stronger memory effect, as discussed in the theoretical analysis subsection.

To detect the scaling behavior of $\mathcal{T}_{\rm eq}$ with respect to $\delta$, we introduce $\mathcal{T}(s, \delta)$, defined by the condition $\langle \Phi(\mathcal{T}, \delta) \rangle = s$, where $s$ serves as a threshold. Figure~\ref{fig-Teq}(c) shows the nonmonotonic relationship between $\mathcal{T}$ and $\delta$ for different values of $s$, corresponding to the three horizontal lines in Fig.~\ref{fig-Teq}(b), from top to bottom. For a small $\delta$, the scaling behavior is independent of $s$, indicating that $\mathcal{T}_{\rm eq} \propto \delta^{-2}$, consistent with the universal thermalization law observed in lattice systems, driven by kinetic effects \cite{PhysRevE.100.010101, Fu_2019, PhysRevE.100.052102, Onorato2019, PhysRevLett.124.186401, PhysRevE.104.L032104, Feng_2022, onorato2023wave, Wang_2024CTP}. Notably, $\mathcal{T}$ reaches its minimum at $\delta = \sqrt{2}$ for short times, as indicated by the black dashed line in Fig.~\ref{fig-Teq}(a). Additionally, Fig.~\ref{fig-Teq}(d) demonstrates that $\mathcal{T}_{\rm eq} \propto (2 - \delta)^{-1/2}$ as $\delta \to 2$. These numerical results strongly support the prediction given by Eq.~(\ref{eq-Teq-delta}).

Figure~\ref{fig-phi}(a) shows the evolution of $\langle \Phi(t) \rangle$ for different system sizes $N$ at the fixed $\delta = 0.05$, exploring the influence of finite-size effects on thermalization behavior. The time is rescaled by $\sqrt{N}$, as it follows from Eq.~(\ref{eq-initial-NT}) that changing the number of particles at a fixed temperature or changing the temperature at a fixed particle number is equivalent in the initial stage. Therefore, the change in system size introduces a time-scale factor similar to the temperature unit. It is evident that $\langle \Phi(t) \rangle$ eventually converges to its theoretical equilibrium value, as indicated by the horizontal dashed lines corresponding to Eq.~(\ref{eq_Phi_theroy}), signifying thermal equilibrium. For small system sizes, the decay follows an exponential form (see the inset), which is nearly independent of system size and rapidly approaches the theoretical value, meaning the system remains in region (I) of Fig.~\ref{fig-taus}.

When the system size exceeds a critical threshold $N_{\rm c}$, a transition occurs from exponential decay to power-law decay, signaling that the system enters region (II). Moreover, multiple oscillations are observed before the system reaches its theoretical value, with peaks occurring around $t \sim N/c_{\rm s}$, marked by the gray vertical line in Fig.~\ref{fig-phi}(b). These oscillations are characteristic of hydrodynamic effects, arising from the collision of acoustic modes under periodic boundary conditions \cite{PhysRevE.89.022111}.

Figure~\ref{fig-phi}(c) shows the results for $\delta = \sqrt{2}$. It is observed that $\langle \Phi(t) \rangle$ exhibits qualitatively identical behavior across all system sizes, decaying in a power-law manner (i.e., $\langle \Phi(t) \rangle\sim t^{-2/3}$) and displaying oscillatory behavior before stabilizing. Furthermore, the onset of oscillations depends on $N$, but becomes independent of $\delta$ for sufficiently large $N$. For example, the curve undergoing power-law decay at $\delta = 0.05$ (indicated by the gray dashed line) closely matches the result obtained at the same size with $\delta = \sqrt{2}$. Additionally, Fig.~\ref{fig-phi}(d) shows nearly complete overlap between all oscillating segments of the curves, suggesting that $\mathcal{T}_{\rm eq} \sim \tau_{\rm h} \simeq N / c_{\rm s}$ (see the gray vertical line). This indicates significant size-dependence due to hydrodynamic effects, specifically the collisions of acoustic modes.

In short, for small $\delta$, when the system size $N$ is less than a critical size $N_{\rm c}$ (i.e., $N < N_{\rm c}$), the kinetic process dominates the thermalization behavior, resulting in an exponential decay of the IPR, with $\mathcal{T}_{\rm eq} \sim \tau_{\rm k} \sim \delta^{-2}$. When $N > N_{\rm c}$, a crossover from exponential to power-law decay occurs due to the predominance of hydrodynamic effects, and $\mathcal{T}_{\rm eq} \sim \tau_{\rm h} \sim N / c_{\rm s}$. As shown in Fig.~\ref{fig-taus}, to observe significant hydrodynamic effects, the condition $\tau_{\rm h} \gg \tau_{\rm k}$ must be met, which implies that $N_{\rm c} \propto \delta^{-2}$, corresponding to the Boltzmann-Grad limit \cite{grad1949kinetic}. Clearly, there are three characteristic regions in the thermalization process of a given system, as illustrated in Fig.~\ref{fig-taus}. For a very small $\delta$, $\langle \Phi(t) \rangle$ decays exponentially, and the system stays in region (I). As $\delta$ increases, $\langle \Phi(t) \rangle$ initially decays exponentially, followed by power-law decay, placing the system in region (II). When $\delta$ approaches $\sqrt{2}$, the decay behavior of $\langle \Phi(t) \rangle$ becomes almost entirely power-law, indicating that the system is in region (III). It is important to note that, in the thermodynamic limit, the order in which the limits are taken can significantly affect the conclusions regarding the thermalization dynamics. For instance, when $N \to \infty$ is taken first, followed by $\delta \to 0$ (observing behavior along the horizontal axis in Fig.\ref{fig-taus}), the integrable limit ensures that the kinetic process dominates. In contrast, if $\delta \to 0$ is taken first, followed by $N \to \infty$ (as along the vertical axis in Fig.~\ref{fig-taus} to observe the relaxation behavior), the system's thermalization dynamics are governed by hydrodynamics. This distinction is crucial in mesoscopic systems, where the order of limits can determine whether the system exhibits symmetry breaking or a cusp catastrophe \cite{QIAN2016153}.

Integrability is an intrinsic property of a system that governs its relaxation dynamics in both nonequilibrium and equilibrium states. According to the Onsager's regression hypothesis \cite{PhysRev.37.405}, the relaxation patterns of nonequilibrium processes and equilibrium fluctuations are identical within the linear response regime. In the following, we examine the relaxation behavior of heat current fluctuations in the DHP system under equilibrium conditions. Based on the linear response theory \cite{kubo1957statistical, kubo1957stat2}, this reflects the system's linear transport properties, allowing us to explore the similarities and differences in the relaxation behaviors of various physical quantities described by the same underlying dynamics in both equilibrium and nonequilibrium states.

\section{Heat Transport}\label{sec:4}

In this section, we examine the heat transport properties of the DHP system by analyzing the heat current autocorrelation function (HCAF) in equilibrium.

\subsection{Theoretical analysis}

The total heat current (HC) is given by $J \equiv \sum_{i=1}^N \frac{m_i v_i^3}{2}$ \cite{casati1986energy}. According to the molecular chaos hypothesis, the HC is a \emph{quasi-stationary quantity}, with its instantaneous rate of change depending solely on its current value \cite{landau2013statistical}. Therefore, the HCAF is expressed as
\begin{equation}\label{eq-Ct1}
  C(t)=\left\langle J(t)J(0)\right\rangle=\left\langle J(0)^2\right\rangle e^{-t/\tau}=C(0)e^{-t/\tau},
\end{equation}
where the Taylor expansion at $t=0$ gives
\begin{equation}\label{eq-Ct2}
  C(t)\approx C(0)-tC(0)/\tau.
\end{equation}
Here, $C(0)=\left\langle J(0)^2\right\rangle$. After a detailed calculation, we obtain
\begin{equation}
C(0)=NT^3\left(\frac{15}{4-\delta^2}-\frac{9}{4}\right).
\end{equation}

\begin{figure*}[t]
  \centering
  \includegraphics[width=2\columnwidth]{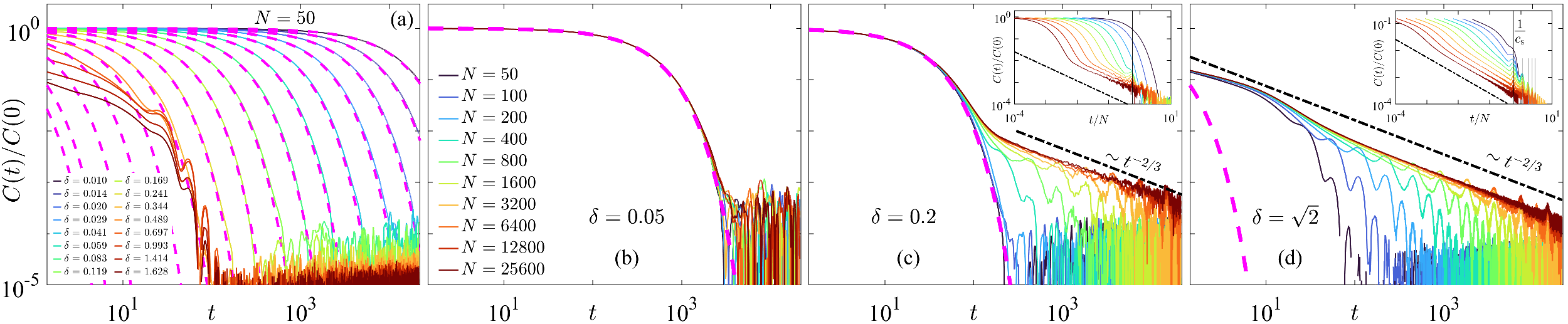}\\
  \caption{(a) Normalized HCAF for various $\delta$ at $N=50$. The magenta dashed lines represent predictions from Eq.~(\ref{eq-Ct1}), with $\tau$ given by Eq.~(\ref{eq-tau}). (b)-(d) Normalized HCAF for different $N$ at $\delta=0.05$, $0.2$, and $\sqrt{2}$, respectively. These panels share the legend. The magenta dashed lines are theoretical predictions. Inset: same as the main panel, but with the horizontal axis rescaled. The black vertical solid lines in the insets correspond to $1/c_{\rm s}$. All black lines with a slope of $-2/3$ are for reference.
  }\label{fig4-corr}
\end{figure*}

Next, considering a collision between particles $i$ and $i+1$, the change in HC due to this collision is
\begin{equation}
\Delta_{1,i}=J_1-J_0=\frac{1}{4}\left(1-\frac{\delta^2}{4}\right)\left(m_i-m_{i+1}\right)\nu_i^3,
\end{equation}
where $J_0$ and $J_1$ represent the HC before and after the collision, and $\nu_i =v_i-v_{i+1}>0$ is the collision condition. The HCAF for a single collision is then
\begin{equation}
C(1) = \left\langle J_1J_0\right\rangle=\left\langle  \left(J_0+\Delta_{1,i}\right)J_0\right\rangle
  =C(0)+\mathcal{S},
\end{equation}
where
\begin{equation}
\mathcal{S}\equiv\left\langle J_0\Delta_{1,i}\right\rangle=-\frac{48T^3 \delta ^2}{4-\delta ^2}.
\end{equation}
For $n$ times collisions, the HCAF becomes
\begin{equation}\label{eq-C(n)}
C(n)=\left\langle J_nJ_0\right\rangle
=C(0)+\xi n \mathcal{S},
\end{equation}
where $\xi={\left\langle J_0\left(\sum\nolimits_{l=1}^n \Delta_{l,i_l}\right)\right\rangle}/{(n\mathcal{S})}$ is a dimensionless coefficient, and $\Delta_{l,i_l}=J_{l}-J_{l-1}$ denotes the change in HC during the $l$th collision between particles $i_l$ and $(i_l+1)$.

Comparing Eqs.~(\ref{eq-Ct2}) and (\ref{eq-C(n)}), and using $t=\frac{n\tau_{\rm f}}{N}$, we obtain
\begin{equation}\label{eq-tauC}
  \tau = -\frac{\tau_{\rm f} C(0)}{\xi N \mathcal{S}}.
\end{equation}
Here, $\xi$ is the sole parameter that needs to be determined. From the thermalization results, we observe that $|\lambda|$ and $g$ can roughly describe the strength of the kinetic and hydrodynamic effects for $\delta\in(0,\sqrt{2})$, where $\delta^2<(4-\delta^2)$, as shown in Fig.~\ref{fig-lambda-g}. Assuming that the kinetic effect vanishes---or, more precisely, that the kinetic and hydrodynamic effects become indistinguishable---when $|\lambda| = g$, the system undergoes a transition from region (II) to region (III) (see Fig.~\ref{fig-taus}), so that $\tau\simeq\tau_{\rm f}$ at $\delta_{\rm c}=\sqrt{2-\sqrt{2}}$. This gives $\xi\approx1/4$, leading to Eq.~(\ref{eq-tauC}) concreted as
\begin{equation}\label{eq-tau}
  \tau =\left(\frac{3}{16}+\frac{1}{2\delta^2}\right)
  \sqrt{\frac{\pi}{T}\left(1-\frac{\delta^2}{4}\right)}.
\end{equation}
Note that as $\delta\to0$, $\tau\propto\delta^{-2}$, which is the same behavior as $\tau_{\rm k}$, as shown in Eq.~(\ref{eq-Teq-delta}). As discussed in Sec.~\ref{sec:2}, when $\delta\to2$, the transport behavior of the system becomes complex, and we will not address it further here. We focus on the range $\delta\in[0,\sqrt{2}]$ going forward.

In addition, the hydrodynamic theory predicts that the HCAF in the large-size limit exhibits a long-time tail, specifically $C(t)\sim t^{-2/3}$ \cite{PhysRevLett.89.200601,PhysRevLett.89.180601,PhysRevLett.108.180601,PhysRevLett.111.230601}. Considering both kinetic and hydrodynamic effects, we decompose the complete HCAF as
\begin{equation}\label{eq_Ckht}
  C(t)=C(0)\left[\eta e^{-t/\tau}+(1-\eta)e^{-1}(t/\tau)^{-2/3}\right],
\end{equation}
where $\eta$ is a weight factor. Specifically, $\eta=1$ for $t<\tau$, and $\eta\in[0,1]$ for $t\geq\tau$.

Following the linear response theory, the heat conductivity can be derived using the Green-Kubo formula as
\begin{equation}\label{eq-GK-kappa}
\kappa = \lim_{t_{\rm c}\to\infty}\lim_{N\to\infty}\frac{1}{T^2N}\int_{0}^{t_{\rm c}}C(t)dt,
\end{equation}
where $t_{\rm c}=N/c_{\rm s}$ is the truncation time \cite{lepri2003thermal}. Inserting Eq.~(\ref{eq-Ct1})
into Eq.~(\ref{eq-GK-kappa}), we obtain the heat conductivity induced by kinetic effects as
\begin{align}\label{eq-kappa-our0}
\kappa_{\rm k} = \frac{(24+9\delta^2)T}{16-4\delta^2}\tau = \frac{3\left(8+3\delta^2\right)^2}{128\delta^2}\sqrt{\frac{\pi T}{4-\delta^2}}
\end{align}
for $t_{\rm c}\gg\tau$. Combining Eq.~(\ref{eq_Ckht})
with Eq.~(\ref{eq-GK-kappa}), we get
\begin{align}\label{eq-kappa-hy}
\kappa(t) = \kappa_{\rm k}\left[\eta(1-e^{-t/\tau})+(1-\eta)h(t)\right],
\end{align}
where
\begin{equation}\label{eq-GK-kappa-hydro}
  h(t)=1-4e^{-1}+3 e^{-1}(t/\tau)^{1/3}
\end{equation}
represents the hydrodynamic contribution. From Eq. (\ref{eq-kappa-our0}), we have $\kappa(\tau)=\frac{3T}{2}\tau$ for $\delta=0$, where the system is integrable and exhibits ballistic transport behavior \cite{kappaForDzeros}.

To address the transport properties in a unified framework, the characteristic time of the HCAF in the kinetic regime, given by Eq.~(\ref{eq-tau}), is shown in Fig.~\ref{fig-taus} (black dashed line). In region (I), the kinetic effects dominate, leading to  exponential decay of the HCAF. In region (II), the three effects coexist, and we will observe a crossover from exponential decay to power-law decay. In region (III), the hydrodynamic effects dominate, resulting in power-law decay of the HCAF.

Furthermore, from Eq.~(\ref{eq-tau}), we see that $\tau / \tau_{\rm f} \propto \delta^{-2}$ as $\delta \to 0$, implying that the kinetic process dominates in the integrable limit, with $\kappa \simeq \kappa_{\rm k} \propto \delta^{-2}$, which is size-independent. To observe significant hydrodynamic effects ($\kappa \gg \kappa_{\rm k}$) for a given $\delta$, the system size $N$ must exceed a critical value $N_{\rm c} \propto \delta^{-2}$, similar to the thermalization process. This requirement indicates that hydrodynamic effects become less noticeable in the integrable limit.
Whether hydrodynamic effects can be observed also depends on the order of taking the limits: $\delta \to 0$ first or $N \to \infty$ first, as discussed in Sec.~\ref{sec:3}. For instance, if $\delta \to 0$ first, hydrodynamic transport behavior may emerge as $N \to \infty$, provided that $N > N_{\rm c} \propto \delta^{-2}$, meaning that $N$ must grow faster than $\delta \to 0$. All of the above analysis will be verified numerically below.

\subsection{Numerical results for heat transport}

Figure~\ref{fig4-corr}(a) shows the normalized HCAF for various values of $\delta$ at the fixed small $N = 50$. For small $\delta$, the kinetic predictions and numerical results agree well. However, as $\delta$ increases, deviations emerge. The kinetic region rapidly shrinks, and hydrodynamic effects begin to dominate, as evidenced by the oscillations arising from the collision of acoustic modes at the boundary \cite{PhysRevE.89.022111}. These results suggest that even small systems exhibit strong hydrodynamic effects when the system is in the far-from-integrable limit, challenging the common belief that hydrodynamic effects are primarily observed in large systems.

Figure~\ref{fig4-corr}(b) presents the normalized HCAF for various system sizes at the fixed $\delta = 0.05$. The HCAF decays exponentially, and all results collapse nearly completely, matching the kinetic predictions (magenta dashed lines). In Fig.~\ref{fig4-corr}(c), for $\delta = 0.2$, the HCAF of small systems still follows an exponential decay and aligns with the kinetic predictions (see the magenta dashed line). However, as the system size increases, a transition from exponential to power-law decay emerges. For larger systems, the power-law decay behavior matches the hydrodynamic prediction, $C(t) \sim t^{-2/3}$ \cite{PhysRevLett.89.200601, PhysRevLett.89.180601, PhysRevLett.108.180601, PhysRevLett.111.230601}.

Figure~\ref{fig4-corr}(d) shows the results for $\delta = \sqrt{2}$, where nearly all curves exhibit power-law decay and hydrodynamic behavior, characterized by oscillations due to the collision of acoustic modes under periodic boundary conditions \cite{PhysRevE.89.022111}. The peaks of the oscillations occur at integer multiples of $N / c_{\rm s}$, as indicated by the vertical lines in the inset. As $N$ increases, the exponent of the power-law decay approaches the theoretical value of $-2/3$ in hydrodynamics. These phenomena are fully consistent with the framework outlined in Fig.~\ref{fig-taus}.

\begin{figure}[t]
  \centering
  \includegraphics[width=1\columnwidth]{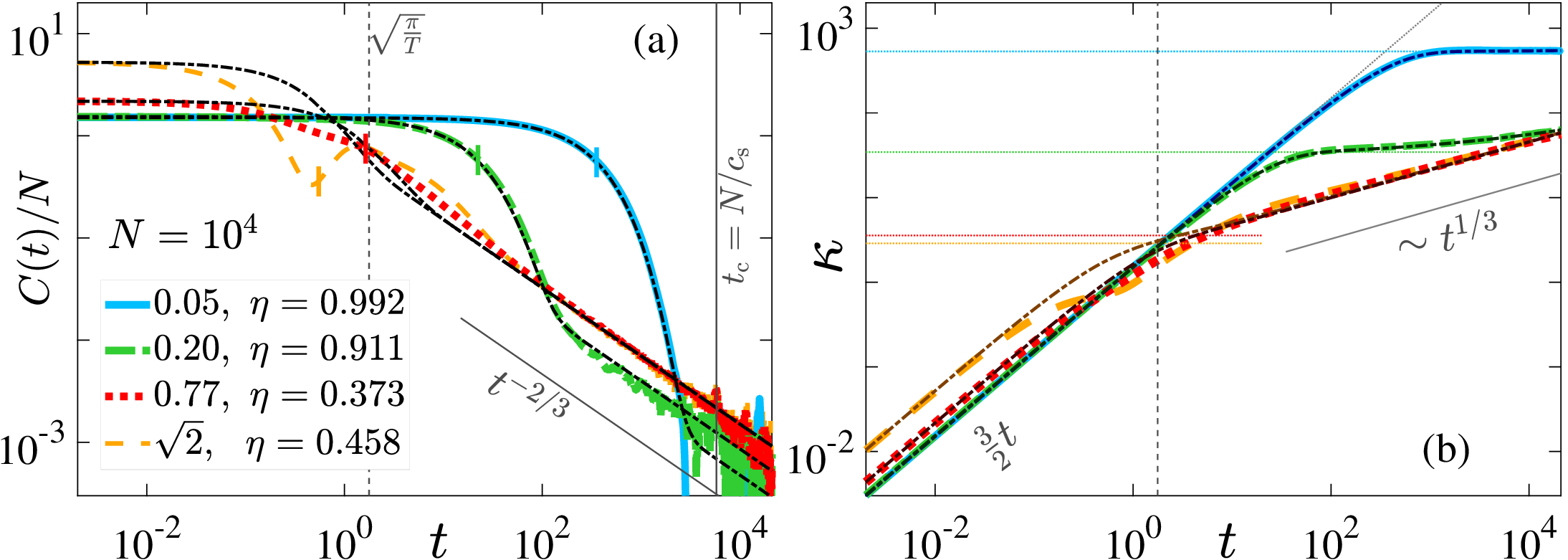}\\
  \caption{(a) Normalized HCAF for various $\delta$. The horizontal coordinate of the short vertical line on the curve indicates the value of $\tau$ from Eq.~(\ref{eq-tau}). The vertical gray dashed line represents the value of $\tau_{\rm f}$ at $\delta=0$, see Eq.~(\ref{eq-meanTime}). The black dashed-dot lines represent the fitting curves using expression (\ref{eq_Ckht}). (b) Heat conductivity $\kappa$ obtained from Eq. (\ref{eq-GK-kappa}). The horizontal lines are predicted from Eq. (\ref{eq-kappa-our0}), and the dark dots represent predictions from Eq. (\ref{eq-kappa-hy}). Both panels share the same legend, and all gray lines serve as reference.
  }\label{fig-corr-kappa}
\end{figure}

Figure~\ref{fig-corr-kappa}(a) shows the HCAF for different values of $\delta$ at a large system size, $N = 10^4$. For all values of $\delta$, over short timescales ($t < \tau_{\rm f}$), the HCAF remains constant, corresponding to a linear increase in heat conductivity ($\kappa$), as shown in Fig.~\ref{fig-corr-kappa}(b), indicating ballistic transport. For $\delta = 0.05$, kinetic processes dominate, and the HCAF decays exponentially, as illustrated by the blue solid line, corresponding to $\kappa \simeq \kappa_{\rm k}$, which indicates normal heat conduction. For $\delta = 0.2$, the exponential decay phase narrows while the power-law decay phase expands, as seen with the green dashed-dotted line. At $\delta = \sqrt{2 - \sqrt{2}} \approx 0.77$, $\tau / \tau_{\rm f} \approx 1$, indicating that the exponential decay phase (kinetic regime) nearly disappears, leading to a direct transition from ballistic to hydrodynamic behavior (power-law decay), as shown by the red dotted line in Fig.~\ref{fig-corr-kappa}(a). This transition is marked by a crossover from $\kappa \sim t$ to $\kappa \sim t^{1/3}$ in Fig.~\ref{fig-corr-kappa}(b). Additionally, oscillations due to acoustic mode collisions are observed, with a characteristic time at $t_{\rm c} \approx 5773.5$ in Fig.~\ref{fig-corr-kappa}(a). As $\delta$ increases further, $\tau < \tau_{\rm f}$, indicating that light particles become localized by heavy particles, which leads to a nonmonotonic decay of the HCAF for $\delta = \sqrt{2}$ (refer to the orange dotted line), with hydrodynamic processes continuing to dominate heat transport.

It is clear that all numerical results are in excellent agreement with theoretical predictions, confirming that the transport properties are entirely determined by the location of system parameters within the phase diagram (see again Fig.~\ref{fig-taus}).
According to the phase diagram, the dependence of the thermal conductivity divergence exponent on the parameter $ \delta $ in the DHP model \cite{hurtado2020simulations} can be naturally understood: different values of $ \delta $ correspond to varying contributions from hydrodynamic effects, which in turn determine the system-size range required to observe the asymptotic scaling law $ \kappa \sim N^{1/3} $. Conversely, for a given system-size range, the $\kappa$-$N$ relationships corresponding to different $\delta$ values exhibit subtle variations, manifesting as distinct effective divergence exponents.

\section{Summary and discussion}\label{sec:5}

In summary, we have investigated the relaxation dynamics of the DHP model, focusing on both thermalization and heat transport. By combining theoretical analysis with high-precision numerical simulations, we have constructed a comprehensive phase diagram that unifies relaxation behavior across regimes ranging from the near-integrable to the far-from-integrable regime.

Our results reveal a qualitatively consistent relaxation structure for both thermalization and heat transport, characterized by three distinct dynamical regimes. In the near-integrable regime, relaxation is dominated by kinetic processes, with both the IPR and HCAF exhibiting exponential decay. The characteristic relaxation time follows the scaling $\tau \propto \delta^{-2}$, and the thermal conductivity $\kappa$ is size-independent ($\kappa \approx \kappa_0$). In the far-from-integrable regime, hydrodynamic processes prevail, with the IPR and HCAF showing power-law decay $\sim t^{-2/3}$. The thermalization time scales linearly with system size ($\tau \propto N$), and $\kappa$ exhibits anomalous scaling as $\kappa \propto N^{1/3}$. The intermediate regime corresponds to the Bogoliubov phase, characterized by a kinetic-to-hydrodynamic crossover and a clear separation of ballistic, kinetic, and hydrodynamic time scales.

The phase diagram clearly delineates how the system's relaxation behavior depends on perturbation strength $\delta$ and system size $N$. Notably, it shows that hydrodynamic behavior can emerge even in small systems when the system is sufficiently far from the integrable regime, challenging the conventional view that such behavior requires the large-size limit. Moreover, we find that  relaxation in the thermodynamic limit depends on the order in which the limits are taken: taking the thermodynamic limit ($N \to \infty$) first yields kinetic-dominated relaxation, while taking the integrable limit ($\delta \to 0$) first results in hydrodynamic dominance. This noncommutativity provides crucial insight into the role of integrability breaking.

The phase diagram can also qualitatively account for the relaxation behavior of 1D lattices. Comparisons with anharmonic chains show consistent patterns: systems with asymmetric IIPs reproduce the DHP scenario, with clear separation of dynamical time scales. Weak nonintegrability (low temperature) yields a sufficiently large kinetic region, leading to normal heat conduction \cite{PhysRevE.85.060102, Zhong_2013, Jiang_2016, Chen_2016}, while strong nonintegrability (high temperature) produces a small kinetic region, resulting in size-dependent hydrodynamic conduction \cite{PhysRevE.88.052112, Das2014}.

This study also sheds light on anharmonic chains with symmetric IIPs. In systems such as the FPUT-$\beta$ chain, the trivial phonon number-conserving resonances dominate the kinetic equation---they are closely linked to strong \emph{phonon hydrodynamic} effects \cite{Ghosh_2022}---which cause frequency shifts but do not effectively promote energy mixing \cite{PhysRevLett.95.264302, Onorato4208, PhysRevLett.120.144301}. Because the dominance of this effect depends primarily on system size rather than the nonintegrability strength, symmetric systems exhibit size-dependent thermalization and heat transport over a broad parameter range. This suggests the need to introduce an additional \emph{phonon hydrodynamic} regime into the phase diagram of symmetric models \cite{Ghosh_2022}, pointing to richer structures that warrant further investigation. Moreover, since the thermal diffusivity satisfies $D = \kappa / c_v$ (with $c_v$ the specific heat \cite{PhysRev.119.1}), diffusion is also expected to follow the same phase-diagram structure.

Overall, this work provides a complete and unified framework linking integrability breaking, thermalization, and heat transport in low-dimensional systems. It offers key insights into relaxation dynamics in quantum systems \cite{RevModPhys.93.025003}, opens new avenues for exploring the interplay between system size, interaction symmetry, and dynamical regimes in both classical and quantum transport phenomena, and highlights the potential of size-independent kinetic transport for the design of high-efficiency thermoelectric materials \cite{PhysRevLett.110.070604, Chen2015, PhysRevLett.121.080602}.

\begin{acknowledgments}
This work was supported by the National Natural Science Foundation of China (Grants Nos. 12465010, 12247106, 12005156, 11975190, 12247101, 12475043, 11925507, and 12047503). W. Fu also acknowledges support from the Youth Talent (Team) Project of Gansu Province (No. 2024QNTD54), the Long-yuan Youth Talents Project of Gansu Province, the Fei-tian Scholars Project of Gansu Province, the Leading Talent Project of Tianshui City, the Innovation Fund from the Department of Education of Gansu Province (Grant No.~2023A-106), the Open Project Program of Key Laboratory of Atomic and Molecular Physics $\&$ Functional Material of Gansu Province (Grant No. 6016-202404), and the Project of Open Competition for the Best Candidates from Department of Education of Gansu Province (Grant No. 2021jyjbgs-06).
\end{acknowledgments}

\bibliography{refgas}

\begin{thebibliography}{122}%
\makeatletter
\providecommand \@ifxundefined [1]{%
 \@ifx{#1\undefined}
}%
\providecommand \@ifnum [1]{%
 \ifnum #1\expandafter \@firstoftwo
 \else \expandafter \@secondoftwo
 \fi
}%
\providecommand \@ifx [1]{%
 \ifx #1\expandafter \@firstoftwo
 \else \expandafter \@secondoftwo
 \fi
}%
\providecommand \natexlab [1]{#1}%
\providecommand \enquote  [1]{``#1''}%
\providecommand \bibnamefont  [1]{#1}%
\providecommand \bibfnamefont [1]{#1}%
\providecommand \citenamefont [1]{#1}%
\providecommand \href@noop [0]{\@secondoftwo}%
\providecommand \href [0]{\begingroup \@sanitize@url \@href}%
\providecommand \@href[1]{\@@startlink{#1}\@@href}%
\providecommand \@@href[1]{\endgroup#1\@@endlink}%
\providecommand \@sanitize@url [0]{\catcode `\\12\catcode `\$12\catcode
  `\&12\catcode `\#12\catcode `\^12\catcode `\_12\catcode `\%12\relax}%
\providecommand \@@startlink[1]{}%
\providecommand \@@endlink[0]{}%
\providecommand \url  [0]{\begingroup\@sanitize@url \@url }%
\providecommand \@url [1]{\endgroup\@href {#1}{\urlprefix }}%
\providecommand \urlprefix  [0]{URL }%
\providecommand \Eprint [0]{\href }%
\providecommand \doibase [0]{https://doi.org/}%
\providecommand \selectlanguage [0]{\@gobble}%
\providecommand \bibinfo  [0]{\@secondoftwo}%
\providecommand \bibfield  [0]{\@secondoftwo}%
\providecommand \translation [1]{[#1]}%
\providecommand \BibitemOpen [0]{}%
\providecommand \bibitemStop [0]{}%
\providecommand \bibitemNoStop [0]{.\EOS\space}%
\providecommand \EOS [0]{\spacefactor3000\relax}%
\providecommand \BibitemShut  [1]{\csname bibitem#1\endcsname}%
\let\auto@bib@innerbib\@empty
%</preamble>
\bibitem [{\citenamefont {Pathria}\ and\ \citenamefont
  {Beale}(2021)}]{pathria2021statistical}%
  \BibitemOpen
  \bibfield  {author} {\bibinfo {author} {\bibfnamefont {R.}~\bibnamefont
  {Pathria}}\ and\ \bibinfo {author} {\bibfnamefont {P.}~\bibnamefont
  {Beale}},\ }\href {https://doi.org/10.1016/C2017-0-01713-5} {\emph {\bibinfo
  {title} {Statistical Mechanics}}},\ \bibinfo {edition} {4th}\ ed.\ (\bibinfo
  {publisher} {Academic Press, Elsevier},\ \bibinfo {year} {2021})\BibitemShut
  {NoStop}%
\bibitem [{\citenamefont {Ford}(1992)}]{FORD1992271}%
  \BibitemOpen
  \bibfield  {author} {\bibinfo {author} {\bibfnamefont {J.}~\bibnamefont
  {Ford}},\ }\bibfield  {title} {\bibinfo {title} {The {Fermi-Pasta-Ulam}
  problem: Paradox turns discovery},\ }\href
  {https://doi.org/https://doi.org/10.1016/0370-1573(92)90116-H} {\bibfield
  {journal} {\bibinfo  {journal} {Phys. Rep.}\ }\textbf {\bibinfo {volume}
  {213}},\ \bibinfo {pages} {271} (\bibinfo {year} {1992})}\BibitemShut
  {NoStop}%
\bibitem [{\citenamefont {Lebowitz}(1993)}]{10.1063/1.881363}%
  \BibitemOpen
  \bibfield  {author} {\bibinfo {author} {\bibfnamefont {J.~L.}\ \bibnamefont
  {Lebowitz}},\ }\bibfield  {title} {\bibinfo {title} {{Boltzmann's Entropy and
  Time's Arrow}},\ }\href {https://doi.org/10.1063/1.881363} {\bibfield
  {journal} {\bibinfo  {journal} {Phys. Today}\ }\textbf {\bibinfo {volume}
  {46}},\ \bibinfo {pages} {32} (\bibinfo {year} {1993})}\BibitemShut {NoStop}%
\bibitem [{\citenamefont {Lebowitz}(1999)}]{RevModPhys.71.S346}%
  \BibitemOpen
  \bibfield  {author} {\bibinfo {author} {\bibfnamefont {J.~L.}\ \bibnamefont
  {Lebowitz}},\ }\bibfield  {title} {\bibinfo {title} {Statistical mechanics: A
  selective review of two central issues},\ }\href
  {https://doi.org/10.1103/RevModPhys.71.S346} {\bibfield  {journal} {\bibinfo
  {journal} {Rev. Mod. Phys.}\ }\textbf {\bibinfo {volume} {71}},\ \bibinfo
  {pages} {S346} (\bibinfo {year} {1999})}\BibitemShut {NoStop}%
\bibitem [{\citenamefont {Fermi}\ \emph {et~al.}(1955)\citenamefont {Fermi},
  \citenamefont {Pasta},\ and\ \citenamefont {Ulam}}]{Fermi1955}%
  \BibitemOpen
  \bibfield  {author} {\bibinfo {author} {\bibfnamefont {E.}~\bibnamefont
  {Fermi}}, \bibinfo {author} {\bibfnamefont {J.}~\bibnamefont {Pasta}},\ and\
  \bibinfo {author} {\bibfnamefont {S.}~\bibnamefont {Ulam}},\ }\bibfield
  {title} {\bibinfo {title} {{Studies of the nonlinear problems}},\ }\href@noop
  {} {\bibfield  {journal} {\bibinfo  {journal} {Los Alamos Scientific
  Laboratory, Report No. LA-1940}\ } (\bibinfo {year} {1955})}\BibitemShut
  {NoStop}%
\bibitem [{\citenamefont {Dauxois}(2008)}]{dauxois:ensl-00202296}%
  \BibitemOpen
  \bibfield  {author} {\bibinfo {author} {\bibfnamefont {T.}~\bibnamefont
  {Dauxois}},\ }\bibfield  {title} {\bibinfo {title} {{Fermi, Pasta, Ulam and a
  mysterious lady}},\ }\href {https://doi.org/10.1063/1.2835154} {\bibfield
  {journal} {\bibinfo  {journal} {{Phys. Today}}\ }\textbf {\bibinfo {volume}
  {61}},\ \bibinfo {pages} {55} (\bibinfo {year} {2008})}\BibitemShut {NoStop}%
\bibitem [{\citenamefont {{Gallavotti}}(2008)}]{2008LNP728G}%
  \BibitemOpen
  \bibinfo {editor} {\bibfnamefont {G.}~\bibnamefont {{Gallavotti}}},\ ed.,\
  \href@noop {} {\emph {\bibinfo {title} {The Fermi-Pasta-Ulam Problem: A
  Status Report}}},\ \bibinfo {series} {Lect. Notes Phys.}, Vol.\ \bibinfo
  {volume} {728}\ (\bibinfo  {publisher} {Springer, New York},\ \bibinfo {year}
  {2008})\BibitemShut {NoStop}%
\bibitem [{\citenamefont {Zabusky}(2005)}]{doi:10.1063/1.1861554}%
  \BibitemOpen
  \bibfield  {author} {\bibinfo {author} {\bibfnamefont {N.~J.}\ \bibnamefont
  {Zabusky}},\ }\bibfield  {title} {\bibinfo {title} {{Fermi-Pasta-Ulam},
  solitons and the fabric of nonlinear and computational science: History,
  synergetics, and visiometrics},\ }\href {https://doi.org/10.1063/1.1861554}
  {\bibfield  {journal} {\bibinfo  {journal} {Chaos}\ }\textbf {\bibinfo
  {volume} {15}},\ \bibinfo {pages} {015102} (\bibinfo {year}
  {2005})}\BibitemShut {NoStop}%
\bibitem [{\citenamefont {Campbell}\ \emph {et~al.}(2005)\citenamefont
  {Campbell}, \citenamefont {Rosenau},\ and\ \citenamefont
  {Zaslavsky}}]{doi:10.1063/1.1889345}%
  \BibitemOpen
  \bibfield  {author} {\bibinfo {author} {\bibfnamefont {D.~K.}\ \bibnamefont
  {Campbell}}, \bibinfo {author} {\bibfnamefont {P.}~\bibnamefont {Rosenau}},\
  and\ \bibinfo {author} {\bibfnamefont {G.~M.}\ \bibnamefont {Zaslavsky}},\
  }\bibfield  {title} {\bibinfo {title} {Introduction: The {Fermi-Pasta-Ulam}
  problem---the first fifty years},\ }\href {https://doi.org/10.1063/1.1889345}
  {\bibfield  {journal} {\bibinfo  {journal} {Chaos}\ }\textbf {\bibinfo
  {volume} {15}},\ \bibinfo {pages} {015101} (\bibinfo {year}
  {2005})}\BibitemShut {NoStop}%
\bibitem [{\citenamefont {Berman}\ and\ \citenamefont
  {Izrailev}(2005)}]{doi:10.1063/1.1855036}%
  \BibitemOpen
  \bibfield  {author} {\bibinfo {author} {\bibfnamefont {G.~P.}\ \bibnamefont
  {Berman}}\ and\ \bibinfo {author} {\bibfnamefont {F.~M.}\ \bibnamefont
  {Izrailev}},\ }\bibfield  {title} {\bibinfo {title} {The {Fermi-Pasta-Ulam}
  problem: Fifty years of progress},\ }\href
  {https://doi.org/10.1063/1.1855036} {\bibfield  {journal} {\bibinfo
  {journal} {Chaos}\ }\textbf {\bibinfo {volume} {15}},\ \bibinfo {pages}
  {015104} (\bibinfo {year} {2005})}\BibitemShut {NoStop}%
\bibitem [{\citenamefont {Pettini}\ \emph {et~al.}(2005)\citenamefont
  {Pettini}, \citenamefont {Casetti}, \citenamefont {Cerruti-Sola},
  \citenamefont {Franzosi},\ and\ \citenamefont
  {Cohen}}]{doi:10.1063/1.1849131}%
  \BibitemOpen
  \bibfield  {author} {\bibinfo {author} {\bibfnamefont {M.}~\bibnamefont
  {Pettini}}, \bibinfo {author} {\bibfnamefont {L.}~\bibnamefont {Casetti}},
  \bibinfo {author} {\bibfnamefont {M.}~\bibnamefont {Cerruti-Sola}}, \bibinfo
  {author} {\bibfnamefont {R.}~\bibnamefont {Franzosi}},\ and\ \bibinfo
  {author} {\bibfnamefont {E.~G.~D.}\ \bibnamefont {Cohen}},\ }\bibfield
  {title} {\bibinfo {title} {Weak and strong chaos in {Fermi-Pasta-Ulam} models
  and beyond},\ }\href {https://doi.org/10.1063/1.1849131} {\bibfield
  {journal} {\bibinfo  {journal} {Chaos}\ }\textbf {\bibinfo {volume} {15}},\
  \bibinfo {pages} {015106} (\bibinfo {year} {2005})}\BibitemShut {NoStop}%
\bibitem [{\citenamefont {Zaslavsky}(2005)}]{doi:10.1063/1.1858115}%
  \BibitemOpen
  \bibfield  {author} {\bibinfo {author} {\bibfnamefont {G.~M.}\ \bibnamefont
  {Zaslavsky}},\ }\bibfield  {title} {\bibinfo {title} {Long way from the
  {FPU}-problem to chaos},\ }\href {https://doi.org/10.1063/1.1858115}
  {\bibfield  {journal} {\bibinfo  {journal} {Chaos}\ }\textbf {\bibinfo
  {volume} {15}},\ \bibinfo {pages} {015103} (\bibinfo {year}
  {2005})}\BibitemShut {NoStop}%
\bibitem [{\citenamefont {Carati}\ \emph {et~al.}(2005)\citenamefont {Carati},
  \citenamefont {Galgani},\ and\ \citenamefont
  {Giorgilli}}]{doi:10.1063/1.1861264}%
  \BibitemOpen
  \bibfield  {author} {\bibinfo {author} {\bibfnamefont {A.}~\bibnamefont
  {Carati}}, \bibinfo {author} {\bibfnamefont {L.}~\bibnamefont {Galgani}},\
  and\ \bibinfo {author} {\bibfnamefont {A.}~\bibnamefont {Giorgilli}},\
  }\bibfield  {title} {\bibinfo {title} {The {Fermi-Pasta-Ulam} problem as a
  challenge for the foundations of physics},\ }\href
  {https://doi.org/10.1063/1.1861264} {\bibfield  {journal} {\bibinfo
  {journal} {Chaos}\ }\textbf {\bibinfo {volume} {15}},\ \bibinfo {pages}
  {015105} (\bibinfo {year} {2005})}\BibitemShut {NoStop}%
\bibitem [{\citenamefont {Fu}\ \emph {et~al.}(2019{\natexlab{a}})\citenamefont
  {Fu}, \citenamefont {Zhang},\ and\ \citenamefont
  {Zhao}}]{PhysRevE.100.010101}%
  \BibitemOpen
  \bibfield  {author} {\bibinfo {author} {\bibfnamefont {W.}~\bibnamefont
  {Fu}}, \bibinfo {author} {\bibfnamefont {Y.}~\bibnamefont {Zhang}},\ and\
  \bibinfo {author} {\bibfnamefont {H.}~\bibnamefont {Zhao}},\ }\bibfield
  {title} {\bibinfo {title} {Universal scaling of the thermalization time in
  one-dimensional lattices},\ }\href
  {https://doi.org/10.1103/PhysRevE.100.010101} {\bibfield  {journal} {\bibinfo
   {journal} {Phys. Rev. E}\ }\textbf {\bibinfo {volume} {100}},\ \bibinfo
  {pages} {010101(R)} (\bibinfo {year} {2019}{\natexlab{a}})}\BibitemShut
  {NoStop}%
\bibitem [{\citenamefont {Fu}\ \emph {et~al.}(2019{\natexlab{b}})\citenamefont
  {Fu}, \citenamefont {Zhang},\ and\ \citenamefont {Zhao}}]{Fu_2019}%
  \BibitemOpen
  \bibfield  {author} {\bibinfo {author} {\bibfnamefont {W.}~\bibnamefont
  {Fu}}, \bibinfo {author} {\bibfnamefont {Y.}~\bibnamefont {Zhang}},\ and\
  \bibinfo {author} {\bibfnamefont {H.}~\bibnamefont {Zhao}},\ }\bibfield
  {title} {\bibinfo {title} {Universal law of thermalization for
  one-dimensional perturbed {Toda} lattices},\ }\href
  {https://doi.org/10.1088/1367-2630/ab115a} {\bibfield  {journal} {\bibinfo
  {journal} {New J. Phys.}\ }\textbf {\bibinfo {volume} {21}},\ \bibinfo
  {pages} {043009} (\bibinfo {year} {2019}{\natexlab{b}})}\BibitemShut
  {NoStop}%
\bibitem [{\citenamefont {Fu}\ \emph {et~al.}(2019{\natexlab{c}})\citenamefont
  {Fu}, \citenamefont {Zhang},\ and\ \citenamefont
  {Zhao}}]{PhysRevE.100.052102}%
  \BibitemOpen
  \bibfield  {author} {\bibinfo {author} {\bibfnamefont {W.}~\bibnamefont
  {Fu}}, \bibinfo {author} {\bibfnamefont {Y.}~\bibnamefont {Zhang}},\ and\
  \bibinfo {author} {\bibfnamefont {H.}~\bibnamefont {Zhao}},\ }\bibfield
  {title} {\bibinfo {title} {Nonintegrability and thermalization of
  one-dimensional diatomic lattices},\ }\href
  {https://doi.org/10.1103/PhysRevE.100.052102} {\bibfield  {journal} {\bibinfo
   {journal} {Phys. Rev. E}\ }\textbf {\bibinfo {volume} {100}},\ \bibinfo
  {pages} {052102} (\bibinfo {year} {2019}{\natexlab{c}})}\BibitemShut
  {NoStop}%
\bibitem [{\citenamefont {Pistone}\ \emph {et~al.}(2019)\citenamefont
  {Pistone}, \citenamefont {Chibbaro}, \citenamefont {Bustamante},
  \citenamefont {Lvov},\ and\ \citenamefont {Onorato}}]{Onorato2019}%
  \BibitemOpen
  \bibfield  {author} {\bibinfo {author} {\bibfnamefont {L.}~\bibnamefont
  {Pistone}}, \bibinfo {author} {\bibfnamefont {S.}~\bibnamefont {Chibbaro}},
  \bibinfo {author} {\bibfnamefont {M.~D.}\ \bibnamefont {Bustamante}},
  \bibinfo {author} {\bibfnamefont {Y.~V.}\ \bibnamefont {Lvov}},\ and\
  \bibinfo {author} {\bibfnamefont {M.}~\bibnamefont {Onorato}},\ }\bibfield
  {title} {\bibinfo {title} {Universal route to thermalization in
  weakly-nonlinear one-dimensional chains},\ }\href
  {https://doi.org/10.3934/mine.2019.4.672} {\bibfield  {journal} {\bibinfo
  {journal} {Math. Eng.}\ }\textbf {\bibinfo {volume} {1}},\ \bibinfo {pages}
  {672} (\bibinfo {year} {2019})}\BibitemShut {NoStop}%
\bibitem [{\citenamefont {Wang}\ \emph {et~al.}(2020)\citenamefont {Wang},
  \citenamefont {Fu}, \citenamefont {Zhang},\ and\ \citenamefont
  {Zhao}}]{PhysRevLett.124.186401}%
  \BibitemOpen
  \bibfield  {author} {\bibinfo {author} {\bibfnamefont {Z.}~\bibnamefont
  {Wang}}, \bibinfo {author} {\bibfnamefont {W.}~\bibnamefont {Fu}}, \bibinfo
  {author} {\bibfnamefont {Y.}~\bibnamefont {Zhang}},\ and\ \bibinfo {author}
  {\bibfnamefont {H.}~\bibnamefont {Zhao}},\ }\bibfield  {title} {\bibinfo
  {title} {Wave-turbulence origin of the instability of {Anderson} localization
  against many-body interactions},\ }\href
  {https://doi.org/10.1103/PhysRevLett.124.186401} {\bibfield  {journal}
  {\bibinfo  {journal} {Phys. Rev. Lett.}\ }\textbf {\bibinfo {volume} {124}},\
  \bibinfo {pages} {186401} (\bibinfo {year} {2020})}\BibitemShut {NoStop}%
\bibitem [{\citenamefont {Fu}\ \emph {et~al.}(2021)\citenamefont {Fu},
  \citenamefont {Zhang},\ and\ \citenamefont {Zhao}}]{PhysRevE.104.L032104}%
  \BibitemOpen
  \bibfield  {author} {\bibinfo {author} {\bibfnamefont {W.}~\bibnamefont
  {Fu}}, \bibinfo {author} {\bibfnamefont {Y.}~\bibnamefont {Zhang}},\ and\
  \bibinfo {author} {\bibfnamefont {H.}~\bibnamefont {Zhao}},\ }\bibfield
  {title} {\bibinfo {title} {Effect of pressure on thermalization of
  one-dimensional nonlinear chains},\ }\href
  {https://doi.org/10.1103/PhysRevE.104.L032104} {\bibfield  {journal}
  {\bibinfo  {journal} {Phys. Rev. E}\ }\textbf {\bibinfo {volume} {104}},\
  \bibinfo {pages} {L032104} (\bibinfo {year} {2021})}\BibitemShut {NoStop}%
\bibitem [{\citenamefont {Feng}\ \emph {et~al.}(2022)\citenamefont {Feng},
  \citenamefont {Fu}, \citenamefont {Zhang},\ and\ \citenamefont
  {Zhao}}]{Feng_2022}%
  \BibitemOpen
  \bibfield  {author} {\bibinfo {author} {\bibfnamefont {S.}~\bibnamefont
  {Feng}}, \bibinfo {author} {\bibfnamefont {W.}~\bibnamefont {Fu}}, \bibinfo
  {author} {\bibfnamefont {Y.}~\bibnamefont {Zhang}},\ and\ \bibinfo {author}
  {\bibfnamefont {H.}~\bibnamefont {Zhao}},\ }\bibfield  {title} {\bibinfo
  {title} {The anti-{Fermi-Pasta-Ulam-Tsingou} problem in one-dimensional
  diatomic lattices},\ }\href {https://doi.org/10.1088/1742-5468/ac6a5a}
  {\bibfield  {journal} {\bibinfo  {journal} {J. Stat. Mech. Theory Exp.}\
  }\textbf {\bibinfo {volume} {2022}},\ \bibinfo {pages} {053104} (\bibinfo
  {year} {2022})}\BibitemShut {NoStop}%
\bibitem [{\citenamefont {Onorato}\ \emph {et~al.}(2023)\citenamefont
  {Onorato}, \citenamefont {Lvov}, \citenamefont {Dematteis},\ and\
  \citenamefont {Chibbaro}}]{onorato2023wave}%
  \BibitemOpen
  \bibfield  {author} {\bibinfo {author} {\bibfnamefont {M.}~\bibnamefont
  {Onorato}}, \bibinfo {author} {\bibfnamefont {Y.~V.}\ \bibnamefont {Lvov}},
  \bibinfo {author} {\bibfnamefont {G.}~\bibnamefont {Dematteis}},\ and\
  \bibinfo {author} {\bibfnamefont {S.}~\bibnamefont {Chibbaro}},\ }\bibfield
  {title} {\bibinfo {title} {Wave turbulence and thermalization in
  one-dimensional chains},\ }\href
  {https://doi.org/10.1016/j.physrep.2023.09.006} {\bibfield  {journal}
  {\bibinfo  {journal} {Phys. Rep.}\ }\textbf {\bibinfo {volume} {1040}},\
  \bibinfo {pages} {1} (\bibinfo {year} {2023})}\BibitemShut {NoStop}%
\bibitem [{\citenamefont {Wang}\ \emph
  {et~al.}(2024{\natexlab{a}})\citenamefont {Wang}, \citenamefont {Fu},
  \citenamefont {Zhang},\ and\ \citenamefont {Zhao}}]{Wang_2024CTP}%
  \BibitemOpen
  \bibfield  {author} {\bibinfo {author} {\bibfnamefont {Z.}~\bibnamefont
  {Wang}}, \bibinfo {author} {\bibfnamefont {W.}~\bibnamefont {Fu}}, \bibinfo
  {author} {\bibfnamefont {Y.}~\bibnamefont {Zhang}},\ and\ \bibinfo {author}
  {\bibfnamefont {H.}~\bibnamefont {Zhao}},\ }\bibfield  {title} {\bibinfo
  {title} {Thermalization of one-dimensional classical lattices: beyond the
  weakly interacting regime},\ }\href
  {https://doi.org/10.1088/1572-9494/ad696d} {\bibfield  {journal} {\bibinfo
  {journal} {Commun. Theor. Phys.}\ }\textbf {\bibinfo {volume} {76}},\
  \bibinfo {pages} {115601} (\bibinfo {year} {2024}{\natexlab{a}})}\BibitemShut
  {NoStop}%
\bibitem [{\citenamefont {Wang}\ \emph
  {et~al.}(2024{\natexlab{b}})\citenamefont {Wang}, \citenamefont {Fu},
  \citenamefont {Zhang},\ and\ \citenamefont {Zhao}}]{Wang24PRL}%
  \BibitemOpen
  \bibfield  {author} {\bibinfo {author} {\bibfnamefont {Z.}~\bibnamefont
  {Wang}}, \bibinfo {author} {\bibfnamefont {W.}~\bibnamefont {Fu}}, \bibinfo
  {author} {\bibfnamefont {Y.}~\bibnamefont {Zhang}},\ and\ \bibinfo {author}
  {\bibfnamefont {H.}~\bibnamefont {Zhao}},\ }\bibfield  {title} {\bibinfo
  {title} {Thermalization of two- and three-dimensional classical lattices},\
  }\href {https://doi.org/10.1103/PhysRevLett.132.217102} {\bibfield  {journal}
  {\bibinfo  {journal} {Phys. Rev. Lett.}\ }\textbf {\bibinfo {volume} {132}},\
  \bibinfo {pages} {217102} (\bibinfo {year} {2024}{\natexlab{b}})}\BibitemShut
  {NoStop}%
\bibitem [{\citenamefont {Danieli}\ \emph {et~al.}(2017)\citenamefont
  {Danieli}, \citenamefont {Campbell},\ and\ \citenamefont
  {Flach}}]{PhysRevE.95.060202}%
  \BibitemOpen
  \bibfield  {author} {\bibinfo {author} {\bibfnamefont {C.}~\bibnamefont
  {Danieli}}, \bibinfo {author} {\bibfnamefont {D.~K.}\ \bibnamefont
  {Campbell}},\ and\ \bibinfo {author} {\bibfnamefont {S.}~\bibnamefont
  {Flach}},\ }\bibfield  {title} {\bibinfo {title} {Intermittent many-body
  dynamics at equilibrium},\ }\href
  {https://doi.org/10.1103/PhysRevE.95.060202} {\bibfield  {journal} {\bibinfo
  {journal} {Phys. Rev. E}\ }\textbf {\bibinfo {volume} {95}},\ \bibinfo
  {pages} {060202} (\bibinfo {year} {2017})}\BibitemShut {NoStop}%
\bibitem [{\citenamefont {Peng}\ \emph {et~al.}(2022)\citenamefont {Peng},
  \citenamefont {Fu}, \citenamefont {Zhang},\ and\ \citenamefont
  {Zhao}}]{Peng_2022}%
  \BibitemOpen
  \bibfield  {author} {\bibinfo {author} {\bibfnamefont {L.}~\bibnamefont
  {Peng}}, \bibinfo {author} {\bibfnamefont {W.}~\bibnamefont {Fu}}, \bibinfo
  {author} {\bibfnamefont {Y.}~\bibnamefont {Zhang}},\ and\ \bibinfo {author}
  {\bibfnamefont {H.}~\bibnamefont {Zhao}},\ }\bibfield  {title} {\bibinfo
  {title} {Instability dynamics of nonlinear normal modes in the
  {Fermi-Pasta-Ulam-Tsingou} chains},\ }\href
  {https://doi.org/10.1088/1367-2630/ac8ac3} {\bibfield  {journal} {\bibinfo
  {journal} {New J. Phys.}\ }\textbf {\bibinfo {volume} {24}},\ \bibinfo
  {pages} {093003} (\bibinfo {year} {2022})}\BibitemShut {NoStop}%
\bibitem [{\citenamefont {Fu}\ \emph {et~al.}(2025)\citenamefont {Fu},
  \citenamefont {Wang}, \citenamefont {Zhang},\ and\ \citenamefont
  {Zhao}}]{fu2025multitype}%
  \BibitemOpen
  \bibfield  {author} {\bibinfo {author} {\bibfnamefont {W.}~\bibnamefont
  {Fu}}, \bibinfo {author} {\bibfnamefont {Z.}~\bibnamefont {Wang}}, \bibinfo
  {author} {\bibfnamefont {Y.}~\bibnamefont {Zhang}},\ and\ \bibinfo {author}
  {\bibfnamefont {H.}~\bibnamefont {Zhao}},\ }\href@noop {} {\bibinfo {title}
  {Multi-type instability processes of periodic orbits in nonlinear chains}}
  (\bibinfo {year} {2025}),\ \Eprint {https://arxiv.org/abs/2501.16821}
  {arXiv:2501.16821 [cond-mat.stat-mech]} \BibitemShut {NoStop}%
\bibitem [{\citenamefont {Lin}\ \emph {et~al.}(2025)\citenamefont {Lin},
  \citenamefont {Fu}, \citenamefont {Wang}, \citenamefont {Zhang},\ and\
  \citenamefont {Zhao}}]{lin2024}%
  \BibitemOpen
  \bibfield  {author} {\bibinfo {author} {\bibfnamefont {W.}~\bibnamefont
  {Lin}}, \bibinfo {author} {\bibfnamefont {W.}~\bibnamefont {Fu}}, \bibinfo
  {author} {\bibfnamefont {Z.}~\bibnamefont {Wang}}, \bibinfo {author}
  {\bibfnamefont {Y.}~\bibnamefont {Zhang}},\ and\ \bibinfo {author}
  {\bibfnamefont {H.}~\bibnamefont {Zhao}},\ }\bibfield  {title} {\bibinfo
  {title} {Universality classes of thermalization and energy diffusion},\
  }\href {https://doi.org/10.1103/PhysRevE.111.024122} {\bibfield  {journal}
  {\bibinfo  {journal} {Phys. Rev. E}\ }\textbf {\bibinfo {volume} {111}},\
  \bibinfo {pages} {024122} (\bibinfo {year} {2025})}\BibitemShut {NoStop}%
\bibitem [{\citenamefont {Langen}\ \emph {et~al.}(2015)\citenamefont {Langen},
  \citenamefont {Geiger},\ and\ \citenamefont {Schmiedmayer}}]{annurev2015}%
  \BibitemOpen
  \bibfield  {author} {\bibinfo {author} {\bibfnamefont {T.}~\bibnamefont
  {Langen}}, \bibinfo {author} {\bibfnamefont {R.}~\bibnamefont {Geiger}},\
  and\ \bibinfo {author} {\bibfnamefont {J.}~\bibnamefont {Schmiedmayer}},\
  }\bibfield  {title} {\bibinfo {title} {Ultracold atoms out of equilibrium},\
  }\href
  {https://doi.org/https://doi.org/10.1146/annurev-conmatphys-031214-014548}
  {\bibfield  {journal} {\bibinfo  {journal} {Annu. Rev. Conden. Ma. P.}\
  }\textbf {\bibinfo {volume} {6}},\ \bibinfo {pages} {201} (\bibinfo {year}
  {2015})}\BibitemShut {NoStop}%
\bibitem [{\citenamefont {Eisert}\ \emph {et~al.}(2015)\citenamefont {Eisert},
  \citenamefont {Friesdorf},\ and\ \citenamefont {Gogolin}}]{Eisert2015}%
  \BibitemOpen
  \bibfield  {author} {\bibinfo {author} {\bibfnamefont {J.}~\bibnamefont
  {Eisert}}, \bibinfo {author} {\bibfnamefont {M.}~\bibnamefont {Friesdorf}},\
  and\ \bibinfo {author} {\bibfnamefont {C.}~\bibnamefont {Gogolin}},\
  }\bibfield  {title} {\bibinfo {title} {Quantum many-body systems out of
  equilibrium},\ }\href {https://doi.org/10.1038/nphys3215} {\bibfield
  {journal} {\bibinfo  {journal} {Nat. Phys.}\ }\textbf {\bibinfo {volume}
  {11}},\ \bibinfo {pages} {124} (\bibinfo {year} {2015})}\BibitemShut
  {NoStop}%
\bibitem [{\citenamefont {Gogolin}\ and\ \citenamefont
  {Eisert}(2016)}]{Gogolin_2016}%
  \BibitemOpen
  \bibfield  {author} {\bibinfo {author} {\bibfnamefont {C.}~\bibnamefont
  {Gogolin}}\ and\ \bibinfo {author} {\bibfnamefont {J.}~\bibnamefont
  {Eisert}},\ }\bibfield  {title} {\bibinfo {title} {Equilibration,
  thermalisation, and the emergence of statistical mechanics in closed quantum
  systems},\ }\href {https://doi.org/10.1088/0034-4885/79/5/056001} {\bibfield
  {journal} {\bibinfo  {journal} {Rep. Prog. Phys.}\ }\textbf {\bibinfo
  {volume} {79}},\ \bibinfo {pages} {056001} (\bibinfo {year}
  {2016})}\BibitemShut {NoStop}%
\bibitem [{\citenamefont {Ueda}(2020)}]{Ueda2020}%
  \BibitemOpen
  \bibfield  {author} {\bibinfo {author} {\bibfnamefont {M.}~\bibnamefont
  {Ueda}},\ }\bibfield  {title} {\bibinfo {title} {Quantum equilibration,
  thermalization and prethermalization in ultracold atoms},\ }\href
  {https://doi.org/10.1038/s42254-020-0237-x} {\bibfield  {journal} {\bibinfo
  {journal} {Nat. Rev. Phys.}\ }\textbf {\bibinfo {volume} {2}},\ \bibinfo
  {pages} {669} (\bibinfo {year} {2020})}\BibitemShut {NoStop}%
\bibitem [{\citenamefont {Rigol}\ \emph {et~al.}(2008)\citenamefont {Rigol},
  \citenamefont {Dunjko},\ and\ \citenamefont {Olshanii}}]{Rigol2008}%
  \BibitemOpen
  \bibfield  {author} {\bibinfo {author} {\bibfnamefont {M.}~\bibnamefont
  {Rigol}}, \bibinfo {author} {\bibfnamefont {V.}~\bibnamefont {Dunjko}},\ and\
  \bibinfo {author} {\bibfnamefont {M.}~\bibnamefont {Olshanii}},\ }\bibfield
  {title} {\bibinfo {title} {Thermalization and its mechanism for generic
  isolated quantum systems},\ }\href {https://doi.org/10.1038/nature06838}
  {\bibfield  {journal} {\bibinfo  {journal} {Nature}\ }\textbf {\bibinfo
  {volume} {452}},\ \bibinfo {pages} {854} (\bibinfo {year}
  {2008})}\BibitemShut {NoStop}%
\bibitem [{\citenamefont {Gring}\ \emph {et~al.}(2012)\citenamefont {Gring},
  \citenamefont {Kuhnert}, \citenamefont {Langen}, \citenamefont {Kitagawa},
  \citenamefont {Rauer}, \citenamefont {Schreitl}, \citenamefont {Mazets},
  \citenamefont {Smith}, \citenamefont {Demler},\ and\ \citenamefont
  {Schmiedmayer}}]{science.1224953}%
  \BibitemOpen
  \bibfield  {author} {\bibinfo {author} {\bibfnamefont {M.}~\bibnamefont
  {Gring}}, \bibinfo {author} {\bibfnamefont {M.}~\bibnamefont {Kuhnert}},
  \bibinfo {author} {\bibfnamefont {T.}~\bibnamefont {Langen}}, \bibinfo
  {author} {\bibfnamefont {T.}~\bibnamefont {Kitagawa}}, \bibinfo {author}
  {\bibfnamefont {B.}~\bibnamefont {Rauer}}, \bibinfo {author} {\bibfnamefont
  {M.}~\bibnamefont {Schreitl}}, \bibinfo {author} {\bibfnamefont
  {I.}~\bibnamefont {Mazets}}, \bibinfo {author} {\bibfnamefont {D.~A.}\
  \bibnamefont {Smith}}, \bibinfo {author} {\bibfnamefont {E.}~\bibnamefont
  {Demler}},\ and\ \bibinfo {author} {\bibfnamefont {J.}~\bibnamefont
  {Schmiedmayer}},\ }\bibfield  {title} {\bibinfo {title} {Relaxation and
  prethermalization in an isolated quantum system},\ }\href
  {https://doi.org/10.1126/science.1224953} {\bibfield  {journal} {\bibinfo
  {journal} {Science}\ }\textbf {\bibinfo {volume} {337}},\ \bibinfo {pages}
  {1318} (\bibinfo {year} {2012})}\BibitemShut {NoStop}%
\bibitem [{\citenamefont {Bastianello}\ \emph {et~al.}(2020)\citenamefont
  {Bastianello}, \citenamefont {De~Luca}, \citenamefont {Doyon},\ and\
  \citenamefont {De~Nardis}}]{PhysRevLett.125.240604}%
  \BibitemOpen
  \bibfield  {author} {\bibinfo {author} {\bibfnamefont {A.}~\bibnamefont
  {Bastianello}}, \bibinfo {author} {\bibfnamefont {A.}~\bibnamefont
  {De~Luca}}, \bibinfo {author} {\bibfnamefont {B.}~\bibnamefont {Doyon}},\
  and\ \bibinfo {author} {\bibfnamefont {J.}~\bibnamefont {De~Nardis}},\
  }\bibfield  {title} {\bibinfo {title} {Thermalization of a trapped
  one-dimensional {Bose} gas via diffusion},\ }\href
  {https://doi.org/10.1103/PhysRevLett.125.240604} {\bibfield  {journal}
  {\bibinfo  {journal} {Phys. Rev. Lett.}\ }\textbf {\bibinfo {volume} {125}},\
  \bibinfo {pages} {240604} (\bibinfo {year} {2020})}\BibitemShut {NoStop}%
\bibitem [{\citenamefont {Korzekwa}\ and\ \citenamefont
  {Lostaglio}(2022)}]{PhysRevLett.129.040602}%
  \BibitemOpen
  \bibfield  {author} {\bibinfo {author} {\bibfnamefont {K.}~\bibnamefont
  {Korzekwa}}\ and\ \bibinfo {author} {\bibfnamefont {M.}~\bibnamefont
  {Lostaglio}},\ }\bibfield  {title} {\bibinfo {title} {Optimizing
  thermalization},\ }\href {https://doi.org/10.1103/PhysRevLett.129.040602}
  {\bibfield  {journal} {\bibinfo  {journal} {Phys. Rev. Lett.}\ }\textbf
  {\bibinfo {volume} {129}},\ \bibinfo {pages} {040602} (\bibinfo {year}
  {2022})}\BibitemShut {NoStop}%
\bibitem [{\citenamefont {Berti}\ \emph {et~al.}(2022)\citenamefont {Berti},
  \citenamefont {Baudin}, \citenamefont {Fusaro}, \citenamefont {Millot},
  \citenamefont {Picozzi},\ and\ \citenamefont
  {Garnier}}]{PhysRevLett.129.063901}%
  \BibitemOpen
  \bibfield  {author} {\bibinfo {author} {\bibfnamefont {N.}~\bibnamefont
  {Berti}}, \bibinfo {author} {\bibfnamefont {K.}~\bibnamefont {Baudin}},
  \bibinfo {author} {\bibfnamefont {A.}~\bibnamefont {Fusaro}}, \bibinfo
  {author} {\bibfnamefont {G.}~\bibnamefont {Millot}}, \bibinfo {author}
  {\bibfnamefont {A.}~\bibnamefont {Picozzi}},\ and\ \bibinfo {author}
  {\bibfnamefont {J.}~\bibnamefont {Garnier}},\ }\bibfield  {title} {\bibinfo
  {title} {Interplay of thermalization and strong disorder: Wave turbulence
  theory, numerical simulations, and experiments in multimode optical fibers},\
  }\href {https://doi.org/10.1103/PhysRevLett.129.063901} {\bibfield  {journal}
  {\bibinfo  {journal} {Phys. Rev. Lett.}\ }\textbf {\bibinfo {volume} {129}},\
  \bibinfo {pages} {063901} (\bibinfo {year} {2022})}\BibitemShut {NoStop}%
\bibitem [{\citenamefont {Darkwah~Oppong}\ \emph {et~al.}(2022)\citenamefont
  {Darkwah~Oppong}, \citenamefont {Pasqualetti}, \citenamefont {Bettermann},
  \citenamefont {Zechmann}, \citenamefont {Knap}, \citenamefont {Bloch},\ and\
  \citenamefont {F\"olling}}]{PhysRevX.12.031026}%
  \BibitemOpen
  \bibfield  {author} {\bibinfo {author} {\bibfnamefont {N.}~\bibnamefont
  {Darkwah~Oppong}}, \bibinfo {author} {\bibfnamefont {G.}~\bibnamefont
  {Pasqualetti}}, \bibinfo {author} {\bibfnamefont {O.}~\bibnamefont
  {Bettermann}}, \bibinfo {author} {\bibfnamefont {P.}~\bibnamefont
  {Zechmann}}, \bibinfo {author} {\bibfnamefont {M.}~\bibnamefont {Knap}},
  \bibinfo {author} {\bibfnamefont {I.}~\bibnamefont {Bloch}},\ and\ \bibinfo
  {author} {\bibfnamefont {S.}~\bibnamefont {F\"olling}},\ }\bibfield  {title}
  {\bibinfo {title} {Probing transport and slow relaxation in the
  mass-imbalanced {Fermi-Hubbard} model},\ }\href
  {https://doi.org/10.1103/PhysRevX.12.031026} {\bibfield  {journal} {\bibinfo
  {journal} {Phys. Rev. X}\ }\textbf {\bibinfo {volume} {12}},\ \bibinfo
  {pages} {031026} (\bibinfo {year} {2022})}\BibitemShut {NoStop}%
\bibitem [{\citenamefont {Le}\ \emph {et~al.}(2023)\citenamefont {Le},
  \citenamefont {Zhang}, \citenamefont {Gopalakrishnan}, \citenamefont
  {Rigol},\ and\ \citenamefont {Weiss}}]{Le2023}%
  \BibitemOpen
  \bibfield  {author} {\bibinfo {author} {\bibfnamefont {Y.}~\bibnamefont
  {Le}}, \bibinfo {author} {\bibfnamefont {Y.}~\bibnamefont {Zhang}}, \bibinfo
  {author} {\bibfnamefont {S.}~\bibnamefont {Gopalakrishnan}}, \bibinfo
  {author} {\bibfnamefont {M.}~\bibnamefont {Rigol}},\ and\ \bibinfo {author}
  {\bibfnamefont {D.~S.}\ \bibnamefont {Weiss}},\ }\bibfield  {title} {\bibinfo
  {title} {Observation of hydrodynamization and local prethermalization in 1d
  {Bose} gases},\ }\href {https://doi.org/10.1038/s41586-023-05979-9}
  {\bibfield  {journal} {\bibinfo  {journal} {Nature}\ }\textbf {\bibinfo
  {volume} {618}},\ \bibinfo {pages} {494} (\bibinfo {year}
  {2023})}\BibitemShut {NoStop}%
\bibitem [{\citenamefont {Garc\'{\i}a~de Soria}\ \emph
  {et~al.}(2024)\citenamefont {Garc\'{\i}a~de Soria}, \citenamefont {Maynar},
  \citenamefont {Gu\'ery-Odelin},\ and\ \citenamefont
  {Trizac}}]{PhysRevLett.132.027101}%
  \BibitemOpen
  \bibfield  {author} {\bibinfo {author} {\bibfnamefont {M.~I.}\ \bibnamefont
  {Garc\'{\i}a~de Soria}}, \bibinfo {author} {\bibfnamefont {P.}~\bibnamefont
  {Maynar}}, \bibinfo {author} {\bibfnamefont {D.}~\bibnamefont
  {Gu\'ery-Odelin}},\ and\ \bibinfo {author} {\bibfnamefont {E.}~\bibnamefont
  {Trizac}},\ }\bibfield  {title} {\bibinfo {title} {Fate of {Boltzmann's}
  breathers: Stokes hypothesis and anomalous thermalization},\ }\href
  {https://doi.org/10.1103/PhysRevLett.132.027101} {\bibfield  {journal}
  {\bibinfo  {journal} {Phys. Rev. Lett.}\ }\textbf {\bibinfo {volume} {132}},\
  \bibinfo {pages} {027101} (\bibinfo {year} {2024})}\BibitemShut {NoStop}%
\bibitem [{\citenamefont {Andersen}\ \emph {et~al.}(2025)\citenamefont
  {Andersen}, \citenamefont {Astrakhantsev}, \citenamefont {Karamlou},\ and\
  \citenamefont {et~al.}}]{Andersen2025}%
  \BibitemOpen
  \bibfield  {author} {\bibinfo {author} {\bibfnamefont {T.~I.}\ \bibnamefont
  {Andersen}}, \bibinfo {author} {\bibfnamefont {N.}~\bibnamefont
  {Astrakhantsev}}, \bibinfo {author} {\bibfnamefont {A.~H.}\ \bibnamefont
  {Karamlou}},\ and\ \bibinfo {author} {\bibnamefont {et~al.}},\ }\bibfield
  {title} {\bibinfo {title} {Thermalization and criticality on an
  analogue--digital quantum simulator},\ }\href
  {https://doi.org/10.1038/s41586-024-08460-3} {\bibfield  {journal} {\bibinfo
  {journal} {Nature}\ }\textbf {\bibinfo {volume} {638}},\ \bibinfo {pages}
  {79} (\bibinfo {year} {2025})}\BibitemShut {NoStop}%
\bibitem [{\citenamefont {Kinoshita}\ \emph {et~al.}(2006)\citenamefont
  {Kinoshita}, \citenamefont {Wenger},\ and\ \citenamefont
  {Weiss}}]{Kinoshita2006}%
  \BibitemOpen
  \bibfield  {author} {\bibinfo {author} {\bibfnamefont {T.}~\bibnamefont
  {Kinoshita}}, \bibinfo {author} {\bibfnamefont {T.}~\bibnamefont {Wenger}},\
  and\ \bibinfo {author} {\bibfnamefont {D.~S.}\ \bibnamefont {Weiss}},\
  }\bibfield  {title} {\bibinfo {title} {A quantum {Newton's} cradle},\ }\href
  {https://doi.org/10.1038/nature04693} {\bibfield  {journal} {\bibinfo
  {journal} {Nature}\ }\textbf {\bibinfo {volume} {440}},\ \bibinfo {pages}
  {900} (\bibinfo {year} {2006})}\BibitemShut {NoStop}%
\bibitem [{\citenamefont {Luca~D'Alessio}\ and\ \citenamefont
  {Rigol}(2016)}]{D'Alessio03052016}%
  \BibitemOpen
  \bibfield  {author} {\bibinfo {author} {\bibfnamefont {A.~P.}\ \bibnamefont
  {Luca~D'Alessio}, \bibfnamefont {Yariv~Kafri}}\ and\ \bibinfo {author}
  {\bibfnamefont {M.}~\bibnamefont {Rigol}},\ }\bibfield  {title} {\bibinfo
  {title} {From quantum chaos and eigenstate thermalization to statistical
  mechanics and thermodynamics},\ }\href
  {https://doi.org/10.1080/00018732.2016.1198134} {\bibfield  {journal}
  {\bibinfo  {journal} {Adv. Phys.}\ }\textbf {\bibinfo {volume} {65}},\
  \bibinfo {pages} {239} (\bibinfo {year} {2016})}\BibitemShut {NoStop}%
\bibitem [{\citenamefont {Mallayya}\ and\ \citenamefont
  {Rigol}(2018)}]{PhysRevLett.120.070603}%
  \BibitemOpen
  \bibfield  {author} {\bibinfo {author} {\bibfnamefont {K.}~\bibnamefont
  {Mallayya}}\ and\ \bibinfo {author} {\bibfnamefont {M.}~\bibnamefont
  {Rigol}},\ }\bibfield  {title} {\bibinfo {title} {Quantum quenches and
  relaxation dynamics in the thermodynamic limit},\ }\href
  {https://doi.org/10.1103/PhysRevLett.120.070603} {\bibfield  {journal}
  {\bibinfo  {journal} {Phys. Rev. Lett.}\ }\textbf {\bibinfo {volume} {120}},\
  \bibinfo {pages} {070603} (\bibinfo {year} {2018})}\BibitemShut {NoStop}%
\bibitem [{\citenamefont {Mallayya}\ \emph {et~al.}(2019)\citenamefont
  {Mallayya}, \citenamefont {Rigol},\ and\ \citenamefont
  {De~Roeck}}]{PhysRevX.9.021027}%
  \BibitemOpen
  \bibfield  {author} {\bibinfo {author} {\bibfnamefont {K.}~\bibnamefont
  {Mallayya}}, \bibinfo {author} {\bibfnamefont {M.}~\bibnamefont {Rigol}},\
  and\ \bibinfo {author} {\bibfnamefont {W.}~\bibnamefont {De~Roeck}},\
  }\bibfield  {title} {\bibinfo {title} {Prethermalization and thermalization
  in isolated quantum systems},\ }\href
  {https://doi.org/10.1103/PhysRevX.9.021027} {\bibfield  {journal} {\bibinfo
  {journal} {Phys. Rev. X}\ }\textbf {\bibinfo {volume} {9}},\ \bibinfo {pages}
  {021027} (\bibinfo {year} {2019})}\BibitemShut {NoStop}%
\bibitem [{\citenamefont {Huveneers}\ and\ \citenamefont
  {Lukkarinen}(2020)}]{PhysRevResearch.2.022034}%
  \BibitemOpen
  \bibfield  {author} {\bibinfo {author} {\bibfnamefont {F.}~\bibnamefont
  {Huveneers}}\ and\ \bibinfo {author} {\bibfnamefont {J.}~\bibnamefont
  {Lukkarinen}},\ }\bibfield  {title} {\bibinfo {title} {Prethermalization in a
  classical phonon field: Slow relaxation of the number of phonons},\ }\href
  {https://doi.org/10.1103/PhysRevResearch.2.022034} {\bibfield  {journal}
  {\bibinfo  {journal} {Phys. Rev. Res.}\ }\textbf {\bibinfo {volume} {2}},\
  \bibinfo {pages} {022034} (\bibinfo {year} {2020})}\BibitemShut {NoStop}%
\bibitem [{\citenamefont {Mallayya}\ and\ \citenamefont
  {Rigol}(2021)}]{PhysRevB.104.184302}%
  \BibitemOpen
  \bibfield  {author} {\bibinfo {author} {\bibfnamefont {K.}~\bibnamefont
  {Mallayya}}\ and\ \bibinfo {author} {\bibfnamefont {M.}~\bibnamefont
  {Rigol}},\ }\bibfield  {title} {\bibinfo {title} {Prethermalization,
  thermalization, and {Fermi's} golden rule in quantum many-body systems},\
  }\href {https://doi.org/10.1103/PhysRevB.104.184302} {\bibfield  {journal}
  {\bibinfo  {journal} {Phys. Rev. B}\ }\textbf {\bibinfo {volume} {104}},\
  \bibinfo {pages} {184302} (\bibinfo {year} {2021})}\BibitemShut {NoStop}%
\bibitem [{\citenamefont {Doyon}\ \emph {et~al.}(2025)\citenamefont {Doyon},
  \citenamefont {Gopalakrishnan}, \citenamefont {M\o{}ller}, \citenamefont
  {Schmiedmayer},\ and\ \citenamefont {Vasseur}}]{PhysRevX.15.010501}%
  \BibitemOpen
  \bibfield  {author} {\bibinfo {author} {\bibfnamefont {B.}~\bibnamefont
  {Doyon}}, \bibinfo {author} {\bibfnamefont {S.}~\bibnamefont
  {Gopalakrishnan}}, \bibinfo {author} {\bibfnamefont {F.}~\bibnamefont
  {M\o{}ller}}, \bibinfo {author} {\bibfnamefont {J.}~\bibnamefont
  {Schmiedmayer}},\ and\ \bibinfo {author} {\bibfnamefont {R.}~\bibnamefont
  {Vasseur}},\ }\bibfield  {title} {\bibinfo {title} {Generalized
  hydrodynamics: A perspective},\ }\href
  {https://doi.org/10.1103/PhysRevX.15.010501} {\bibfield  {journal} {\bibinfo
  {journal} {Phys. Rev. X}\ }\textbf {\bibinfo {volume} {15}},\ \bibinfo
  {pages} {010501} (\bibinfo {year} {2025})}\BibitemShut {NoStop}%
\bibitem [{\citenamefont {Shastry}\ and\ \citenamefont
  {Young}(2010)}]{PhysRevB.82.104306}%
  \BibitemOpen
  \bibfield  {author} {\bibinfo {author} {\bibfnamefont {B.~S.}\ \bibnamefont
  {Shastry}}\ and\ \bibinfo {author} {\bibfnamefont {A.~P.}\ \bibnamefont
  {Young}},\ }\bibfield  {title} {\bibinfo {title} {Dynamics of energy
  transport in a {Toda} ring},\ }\href
  {https://doi.org/10.1103/PhysRevB.82.104306} {\bibfield  {journal} {\bibinfo
  {journal} {Phys. Rev. B}\ }\textbf {\bibinfo {volume} {82}},\ \bibinfo
  {pages} {104306} (\bibinfo {year} {2010})}\BibitemShut {NoStop}%
\bibitem [{\citenamefont {Mori}\ \emph {et~al.}(2018)\citenamefont {Mori},
  \citenamefont {Ikeda}, \citenamefont {Kaminishi},\ and\ \citenamefont
  {Ueda}}]{Mori_2018}%
  \BibitemOpen
  \bibfield  {author} {\bibinfo {author} {\bibfnamefont {T.}~\bibnamefont
  {Mori}}, \bibinfo {author} {\bibfnamefont {T.~N.}\ \bibnamefont {Ikeda}},
  \bibinfo {author} {\bibfnamefont {E.}~\bibnamefont {Kaminishi}},\ and\
  \bibinfo {author} {\bibfnamefont {M.}~\bibnamefont {Ueda}},\ }\bibfield
  {title} {\bibinfo {title} {Thermalization and prethermalization in isolated
  quantum systems: a theoretical overview},\ }\href
  {https://doi.org/10.1088/1361-6455/aabcdf} {\bibfield  {journal} {\bibinfo
  {journal} {J. Phys. B: At. Mol. Opt. Phys.}\ }\textbf {\bibinfo {volume}
  {51}},\ \bibinfo {pages} {112001} (\bibinfo {year} {2018})}\BibitemShut
  {NoStop}%
\bibitem [{\citenamefont {Gogolin}\ \emph {et~al.}(2011)\citenamefont
  {Gogolin}, \citenamefont {M\"uller},\ and\ \citenamefont
  {Eisert}}]{PhysRevLett.106.040401}%
  \BibitemOpen
  \bibfield  {author} {\bibinfo {author} {\bibfnamefont {C.}~\bibnamefont
  {Gogolin}}, \bibinfo {author} {\bibfnamefont {M.~P.}\ \bibnamefont
  {M\"uller}},\ and\ \bibinfo {author} {\bibfnamefont {J.}~\bibnamefont
  {Eisert}},\ }\bibfield  {title} {\bibinfo {title} {Absence of thermalization
  in nonintegrable systems},\ }\href
  {https://doi.org/10.1103/PhysRevLett.106.040401} {\bibfield  {journal}
  {\bibinfo  {journal} {Phys. Rev. Lett.}\ }\textbf {\bibinfo {volume} {106}},\
  \bibinfo {pages} {040401} (\bibinfo {year} {2011})}\BibitemShut {NoStop}%
\bibitem [{\citenamefont {Balz}\ and\ \citenamefont
  {Reimann}(2017)}]{PhysRevLett.118.190601}%
  \BibitemOpen
  \bibfield  {author} {\bibinfo {author} {\bibfnamefont {B.~N.}\ \bibnamefont
  {Balz}}\ and\ \bibinfo {author} {\bibfnamefont {P.}~\bibnamefont {Reimann}},\
  }\bibfield  {title} {\bibinfo {title} {Typical relaxation of isolated
  many-body systems which do not thermalize},\ }\href
  {https://doi.org/10.1103/PhysRevLett.118.190601} {\bibfield  {journal}
  {\bibinfo  {journal} {Phys. Rev. Lett.}\ }\textbf {\bibinfo {volume} {118}},\
  \bibinfo {pages} {190601} (\bibinfo {year} {2017})}\BibitemShut {NoStop}%
\bibitem [{\citenamefont {Durnin}\ \emph {et~al.}(2021)\citenamefont {Durnin},
  \citenamefont {Bhaseen},\ and\ \citenamefont
  {Doyon}}]{PhysRevLett.127.130601}%
  \BibitemOpen
  \bibfield  {author} {\bibinfo {author} {\bibfnamefont {J.}~\bibnamefont
  {Durnin}}, \bibinfo {author} {\bibfnamefont {M.~J.}\ \bibnamefont
  {Bhaseen}},\ and\ \bibinfo {author} {\bibfnamefont {B.}~\bibnamefont
  {Doyon}},\ }\bibfield  {title} {\bibinfo {title} {Nonequilibrium dynamics and
  weakly broken integrability},\ }\href
  {https://doi.org/10.1103/PhysRevLett.127.130601} {\bibfield  {journal}
  {\bibinfo  {journal} {Phys. Rev. Lett.}\ }\textbf {\bibinfo {volume} {127}},\
  \bibinfo {pages} {130601} (\bibinfo {year} {2021})}\BibitemShut {NoStop}%
\bibitem [{\citenamefont {Abanin}\ \emph {et~al.}(2017)\citenamefont {Abanin},
  \citenamefont {De~Roeck}, \citenamefont {Ho},\ and\ \citenamefont
  {Huveneers}}]{Abanin2017}%
  \BibitemOpen
  \bibfield  {author} {\bibinfo {author} {\bibfnamefont {D.}~\bibnamefont
  {Abanin}}, \bibinfo {author} {\bibfnamefont {W.}~\bibnamefont {De~Roeck}},
  \bibinfo {author} {\bibfnamefont {W.~W.}\ \bibnamefont {Ho}},\ and\ \bibinfo
  {author} {\bibfnamefont {F.}~\bibnamefont {Huveneers}},\ }\bibfield  {title}
  {\bibinfo {title} {A rigorous theory of many-body prethermalization for
  periodically driven and closed quantum systems},\ }\href
  {https://doi.org/10.1007/s00220-017-2930-x} {\bibfield  {journal} {\bibinfo
  {journal} {Commun. Math. Phys.}\ }\textbf {\bibinfo {volume} {354}},\
  \bibinfo {pages} {809} (\bibinfo {year} {2017})}\BibitemShut {NoStop}%
\bibitem [{\citenamefont {Surace}\ and\ \citenamefont
  {Motrunich}(2023)}]{PhysRevResearch.5.043019}%
  \BibitemOpen
  \bibfield  {author} {\bibinfo {author} {\bibfnamefont {F.~M.}\ \bibnamefont
  {Surace}}\ and\ \bibinfo {author} {\bibfnamefont {O.}~\bibnamefont
  {Motrunich}},\ }\bibfield  {title} {\bibinfo {title} {Weak integrability
  breaking perturbations of integrable models},\ }\href
  {https://doi.org/10.1103/PhysRevResearch.5.043019} {\bibfield  {journal}
  {\bibinfo  {journal} {Phys. Rev. Res.}\ }\textbf {\bibinfo {volume} {5}},\
  \bibinfo {pages} {043019} (\bibinfo {year} {2023})}\BibitemShut {NoStop}%
\bibitem [{\citenamefont {Kubo}(1966)}]{R_Kubo_1966}%
  \BibitemOpen
  \bibfield  {author} {\bibinfo {author} {\bibfnamefont {R.}~\bibnamefont
  {Kubo}},\ }\bibfield  {title} {\bibinfo {title} {The fluctuation-dissipation
  theorem},\ }\href {https://doi.org/10.1088/0034-4885/29/1/306} {\bibfield
  {journal} {\bibinfo  {journal} {Rep. Prog. Phys.}\ }\textbf {\bibinfo
  {volume} {29}},\ \bibinfo {pages} {255} (\bibinfo {year} {1966})}\BibitemShut
  {NoStop}%
\bibitem [{\citenamefont {Lepri}\ \emph {et~al.}(2003)\citenamefont {Lepri},
  \citenamefont {Livi},\ and\ \citenamefont {Politi}}]{lepri2003thermal}%
  \BibitemOpen
  \bibfield  {author} {\bibinfo {author} {\bibfnamefont {S.}~\bibnamefont
  {Lepri}}, \bibinfo {author} {\bibfnamefont {R.}~\bibnamefont {Livi}},\ and\
  \bibinfo {author} {\bibfnamefont {A.}~\bibnamefont {Politi}},\ }\bibfield
  {title} {\bibinfo {title} {Thermal conduction in classical low-dimensional
  lattices},\ }\href {https://doi.org/10.1016/S0370-1573(02)00558-6} {\bibfield
   {journal} {\bibinfo  {journal} {Phys. Rep.}\ }\textbf {\bibinfo {volume}
  {377}},\ \bibinfo {pages} {1} (\bibinfo {year} {2003})}\BibitemShut {NoStop}%
\bibitem [{\citenamefont {Dhar}(2008)}]{Dhar08AdvPhys}%
  \BibitemOpen
  \bibfield  {author} {\bibinfo {author} {\bibfnamefont {A.}~\bibnamefont
  {Dhar}},\ }\bibfield  {title} {\bibinfo {title} {Heat transport in
  low-dimensional systems},\ }\href {https://doi.org/10.1080/00018730802538522}
  {\bibfield  {journal} {\bibinfo  {journal} {Adv Phys}\ }\textbf {\bibinfo
  {volume} {57}},\ \bibinfo {pages} {457} (\bibinfo {year} {2008})}\BibitemShut
  {NoStop}%
\bibitem [{\citenamefont {Lepri}(2016)}]{Lepri2016}%
  \BibitemOpen
  \bibfield  {author} {\bibinfo {author} {\bibfnamefont {S.}~\bibnamefont
  {Lepri}},\ }\href {https://doi.org/10.1007/978-3-319-29261-8} {\emph
  {\bibinfo {title} {Thermal Transport in Low Dimensions: From Statistical
  Physics to Nanoscale Heat Transfer}}},\ Lecture Notes in Physics\ (\bibinfo
  {publisher} {Springer Cham},\ \bibinfo {year} {2016})\BibitemShut {NoStop}%
\bibitem [{\citenamefont {Narayan}\ and\ \citenamefont
  {Ramaswamy}(2002)}]{PhysRevLett.89.200601}%
  \BibitemOpen
  \bibfield  {author} {\bibinfo {author} {\bibfnamefont {O.}~\bibnamefont
  {Narayan}}\ and\ \bibinfo {author} {\bibfnamefont {S.}~\bibnamefont
  {Ramaswamy}},\ }\bibfield  {title} {\bibinfo {title} {Anomalous heat
  conduction in one-dimensional momentum-conserving systems},\ }\href
  {https://doi.org/10.1103/PhysRevLett.89.200601} {\bibfield  {journal}
  {\bibinfo  {journal} {Phys. Rev. Lett.}\ }\textbf {\bibinfo {volume} {89}},\
  \bibinfo {pages} {200601} (\bibinfo {year} {2002})}\BibitemShut {NoStop}%
\bibitem [{\citenamefont {van Beijeren}(2012)}]{PhysRevLett.108.180601}%
  \BibitemOpen
  \bibfield  {author} {\bibinfo {author} {\bibfnamefont {H.}~\bibnamefont {van
  Beijeren}},\ }\bibfield  {title} {\bibinfo {title} {Exact results for
  anomalous transport in one-dimensional hamiltonian systems},\ }\href
  {https://doi.org/10.1103/PhysRevLett.108.180601} {\bibfield  {journal}
  {\bibinfo  {journal} {Phys. Rev. Lett.}\ }\textbf {\bibinfo {volume} {108}},\
  \bibinfo {pages} {180601} (\bibinfo {year} {2012})}\BibitemShut {NoStop}%
\bibitem [{\citenamefont {Mendl}\ and\ \citenamefont
  {Spohn}(2013)}]{PhysRevLett.111.230601}%
  \BibitemOpen
  \bibfield  {author} {\bibinfo {author} {\bibfnamefont {C.~B.}\ \bibnamefont
  {Mendl}}\ and\ \bibinfo {author} {\bibfnamefont {H.}~\bibnamefont {Spohn}},\
  }\bibfield  {title} {\bibinfo {title} {Dynamic correlators of
  fermi-pasta-ulam chains and nonlinear fluctuating hydrodynamics},\ }\href
  {https://doi.org/10.1103/PhysRevLett.111.230601} {\bibfield  {journal}
  {\bibinfo  {journal} {Phys. Rev. Lett.}\ }\textbf {\bibinfo {volume} {111}},\
  \bibinfo {pages} {230601} (\bibinfo {year} {2013})}\BibitemShut {NoStop}%
\bibitem [{\citenamefont {Spohn}(2014)}]{Spohn2014}%
  \BibitemOpen
  \bibfield  {author} {\bibinfo {author} {\bibfnamefont {H.}~\bibnamefont
  {Spohn}},\ }\bibfield  {title} {\bibinfo {title} {Nonlinear fluctuating
  hydrodynamics for anharmonic chains},\ }\href
  {https://doi.org/10.1007/s10955-014-0933-y} {\bibfield  {journal} {\bibinfo
  {journal} {J. Stat. Phys.}\ }\textbf {\bibinfo {volume} {154}},\ \bibinfo
  {pages} {1191} (\bibinfo {year} {2014})}\BibitemShut {NoStop}%
\bibitem [{\citenamefont {Delfini}\ \emph {et~al.}(2006)\citenamefont
  {Delfini}, \citenamefont {Lepri}, \citenamefont {Livi},\ and\ \citenamefont
  {Politi}}]{PhysRevE.73.060201}%
  \BibitemOpen
  \bibfield  {author} {\bibinfo {author} {\bibfnamefont {L.}~\bibnamefont
  {Delfini}}, \bibinfo {author} {\bibfnamefont {S.}~\bibnamefont {Lepri}},
  \bibinfo {author} {\bibfnamefont {R.}~\bibnamefont {Livi}},\ and\ \bibinfo
  {author} {\bibfnamefont {A.}~\bibnamefont {Politi}},\ }\bibfield  {title}
  {\bibinfo {title} {Self-consistent mode-coupling approach to one-dimensional
  heat transport},\ }\href {https://doi.org/10.1103/PhysRevE.73.060201}
  {\bibfield  {journal} {\bibinfo  {journal} {Phys. Rev. E}\ }\textbf {\bibinfo
  {volume} {73}},\ \bibinfo {pages} {060201} (\bibinfo {year}
  {2006})}\BibitemShut {NoStop}%
\bibitem [{\citenamefont {Delfini}\ \emph {et~al.}(2007)\citenamefont
  {Delfini}, \citenamefont {Lepri}, \citenamefont {Livi},\ and\ \citenamefont
  {Politi}}]{Delfini_2007}%
  \BibitemOpen
  \bibfield  {author} {\bibinfo {author} {\bibfnamefont {L.}~\bibnamefont
  {Delfini}}, \bibinfo {author} {\bibfnamefont {S.}~\bibnamefont {Lepri}},
  \bibinfo {author} {\bibfnamefont {R.}~\bibnamefont {Livi}},\ and\ \bibinfo
  {author} {\bibfnamefont {A.}~\bibnamefont {Politi}},\ }\bibfield  {title}
  {\bibinfo {title} {Anomalous kinetics and transport from 1d self-consistent
  mode-coupling theory},\ }\href
  {https://doi.org/10.1088/1742-5468/2007/02/P02007} {\bibfield  {journal}
  {\bibinfo  {journal} {J. Stat. Mech: Theory Exp.}\ }\textbf {\bibinfo
  {volume} {2007}},\ \bibinfo {pages} {P02007} (\bibinfo {year}
  {2007})}\BibitemShut {NoStop}%
\bibitem [{\citenamefont {Chen}\ \emph
  {et~al.}(2014{\natexlab{a}})\citenamefont {Chen}, \citenamefont {Zhang},
  \citenamefont {Wang},\ and\ \citenamefont {Zhao}}]{PhysRevE.89.022111}%
  \BibitemOpen
  \bibfield  {author} {\bibinfo {author} {\bibfnamefont {S.}~\bibnamefont
  {Chen}}, \bibinfo {author} {\bibfnamefont {Y.}~\bibnamefont {Zhang}},
  \bibinfo {author} {\bibfnamefont {J.}~\bibnamefont {Wang}},\ and\ \bibinfo
  {author} {\bibfnamefont {H.}~\bibnamefont {Zhao}},\ }\bibfield  {title}
  {\bibinfo {title} {Finite-size effects on current correlation functions},\
  }\href {https://doi.org/10.1103/PhysRevE.89.022111} {\bibfield  {journal}
  {\bibinfo  {journal} {Phys. Rev. E}\ }\textbf {\bibinfo {volume} {89}},\
  \bibinfo {pages} {022111} (\bibinfo {year} {2014}{\natexlab{a}})}\BibitemShut
  {NoStop}%
\bibitem [{\citenamefont {Lepri}\ \emph {et~al.}(1998)\citenamefont {Lepri},
  \citenamefont {Livi},\ and\ \citenamefont {Politi}}]{S.Lepri_1998}%
  \BibitemOpen
  \bibfield  {author} {\bibinfo {author} {\bibfnamefont {S.}~\bibnamefont
  {Lepri}}, \bibinfo {author} {\bibfnamefont {R.}~\bibnamefont {Livi}},\ and\
  \bibinfo {author} {\bibfnamefont {A.}~\bibnamefont {Politi}},\ }\bibfield
  {title} {\bibinfo {title} {On the anomalous thermal conductivity of
  one-dimensional lattices},\ }\href
  {https://doi.org/10.1209/epl/i1998-00352-3} {\bibfield  {journal} {\bibinfo
  {journal} {Europhys. Lett.}\ }\textbf {\bibinfo {volume} {43}},\ \bibinfo
  {pages} {271} (\bibinfo {year} {1998})}\BibitemShut {NoStop}%
\bibitem [{\citenamefont {Lepri}(1998)}]{PhysRevE.58.7165}%
  \BibitemOpen
  \bibfield  {author} {\bibinfo {author} {\bibfnamefont {S.}~\bibnamefont
  {Lepri}},\ }\bibfield  {title} {\bibinfo {title} {Relaxation of classical
  many-body hamiltonians in one dimension},\ }\href
  {https://doi.org/10.1103/PhysRevE.58.7165} {\bibfield  {journal} {\bibinfo
  {journal} {Phys. Rev. E}\ }\textbf {\bibinfo {volume} {58}},\ \bibinfo
  {pages} {7165} (\bibinfo {year} {1998})}\BibitemShut {NoStop}%
\bibitem [{\citenamefont {Wang}\ and\ \citenamefont
  {Li}(2004)}]{PhysRevLett.92.074302}%
  \BibitemOpen
  \bibfield  {author} {\bibinfo {author} {\bibfnamefont {J.-S.}\ \bibnamefont
  {Wang}}\ and\ \bibinfo {author} {\bibfnamefont {B.}~\bibnamefont {Li}},\
  }\bibfield  {title} {\bibinfo {title} {Intriguing heat conduction of a chain
  with transverse motions},\ }\href
  {https://doi.org/10.1103/PhysRevLett.92.074302} {\bibfield  {journal}
  {\bibinfo  {journal} {Phys. Rev. Lett.}\ }\textbf {\bibinfo {volume} {92}},\
  \bibinfo {pages} {074302} (\bibinfo {year} {2004})}\BibitemShut {NoStop}%
\bibitem [{\citenamefont {Pereverzev}(2003)}]{PhysRevE.68.056124}%
  \BibitemOpen
  \bibfield  {author} {\bibinfo {author} {\bibfnamefont {A.}~\bibnamefont
  {Pereverzev}},\ }\bibfield  {title} {\bibinfo {title} {Fermi-pasta-ulam
  $\ensuremath{\beta}$ lattice: Peierls equation and anomalous heat
  conductivity},\ }\href {https://doi.org/10.1103/PhysRevE.68.056124}
  {\bibfield  {journal} {\bibinfo  {journal} {Phys. Rev. E}\ }\textbf {\bibinfo
  {volume} {68}},\ \bibinfo {pages} {056124} (\bibinfo {year}
  {2003})}\BibitemShut {NoStop}%
\bibitem [{\citenamefont {Dematteis}\ \emph {et~al.}(2020)\citenamefont
  {Dematteis}, \citenamefont {Rondoni}, \citenamefont {Proment}, \citenamefont
  {De~Vita},\ and\ \citenamefont {Onorato}}]{PhysRevLett.125.024101}%
  \BibitemOpen
  \bibfield  {author} {\bibinfo {author} {\bibfnamefont {G.}~\bibnamefont
  {Dematteis}}, \bibinfo {author} {\bibfnamefont {L.}~\bibnamefont {Rondoni}},
  \bibinfo {author} {\bibfnamefont {D.}~\bibnamefont {Proment}}, \bibinfo
  {author} {\bibfnamefont {F.}~\bibnamefont {De~Vita}},\ and\ \bibinfo {author}
  {\bibfnamefont {M.}~\bibnamefont {Onorato}},\ }\bibfield  {title} {\bibinfo
  {title} {Coexistence of ballistic and fourier regimes in the
  $\ensuremath{\beta}$ fermi-pasta-ulam-tsingou lattice},\ }\href
  {https://doi.org/10.1103/PhysRevLett.125.024101} {\bibfield  {journal}
  {\bibinfo  {journal} {Phys. Rev. Lett.}\ }\textbf {\bibinfo {volume} {125}},\
  \bibinfo {pages} {024101} (\bibinfo {year} {2020})}\BibitemShut {NoStop}%
\bibitem [{\citenamefont {Takatsu}\ \emph {et~al.}(2024)\citenamefont
  {Takatsu}, \citenamefont {Kitamura},\ and\ \citenamefont
  {Yoshimura}}]{2024JPSJ.93.053001}%
  \BibitemOpen
  \bibfield  {author} {\bibinfo {author} {\bibfnamefont {M.}~\bibnamefont
  {Takatsu}}, \bibinfo {author} {\bibfnamefont {T.}~\bibnamefont {Kitamura}},\
  and\ \bibinfo {author} {\bibfnamefont {K.}~\bibnamefont {Yoshimura}},\
  }\bibfield  {title} {\bibinfo {title} {A large scale non-equilibrium
  simulation of heat transport in one-dimensional
  {Fermi-Pasta–Ulam–Tsingou} lattice},\ }\href
  {https://doi.org/10.7566/JPSJ.93.053001} {\bibfield  {journal} {\bibinfo
  {journal} {J. Phys. Soc. Jpn.}\ }\textbf {\bibinfo {volume} {93}},\ \bibinfo
  {pages} {053001} (\bibinfo {year} {2024})}\BibitemShut {NoStop}%
\bibitem [{\citenamefont {Zhong}\ \emph {et~al.}(2012)\citenamefont {Zhong},
  \citenamefont {Zhang}, \citenamefont {Wang},\ and\ \citenamefont
  {Zhao}}]{PhysRevE.85.060102}%
  \BibitemOpen
  \bibfield  {author} {\bibinfo {author} {\bibfnamefont {Y.}~\bibnamefont
  {Zhong}}, \bibinfo {author} {\bibfnamefont {Y.}~\bibnamefont {Zhang}},
  \bibinfo {author} {\bibfnamefont {J.}~\bibnamefont {Wang}},\ and\ \bibinfo
  {author} {\bibfnamefont {H.}~\bibnamefont {Zhao}},\ }\bibfield  {title}
  {\bibinfo {title} {Normal heat conduction in one-dimensional momentum
  conserving lattices with asymmetric interactions},\ }\href
  {https://doi.org/10.1103/PhysRevE.85.060102} {\bibfield  {journal} {\bibinfo
  {journal} {Phys. Rev. E}\ }\textbf {\bibinfo {volume} {85}},\ \bibinfo
  {pages} {060102(R)} (\bibinfo {year} {2012})}\BibitemShut {NoStop}%
\bibitem [{\citenamefont {Zhong}\ \emph {et~al.}(2013)\citenamefont {Zhong},
  \citenamefont {Zhang}, \citenamefont {Wang},\ and\ \citenamefont
  {Zhao}}]{Zhong_2013}%
  \BibitemOpen
  \bibfield  {author} {\bibinfo {author} {\bibfnamefont {Y.}~\bibnamefont
  {Zhong}}, \bibinfo {author} {\bibfnamefont {Y.}~\bibnamefont {Zhang}},
  \bibinfo {author} {\bibfnamefont {J.}~\bibnamefont {Wang}},\ and\ \bibinfo
  {author} {\bibfnamefont {H.}~\bibnamefont {Zhao}},\ }\bibfield  {title}
  {\bibinfo {title} {Normal thermal conduction in lattice models with
  asymmetric harmonic interparticle interactions},\ }\href
  {https://doi.org/10.1088/1674-1056/22/7/070505} {\bibfield  {journal}
  {\bibinfo  {journal} {Chinese Phys. B}\ }\textbf {\bibinfo {volume} {22}},\
  \bibinfo {pages} {070505} (\bibinfo {year} {2013})}\BibitemShut {NoStop}%
\bibitem [{\citenamefont {Chen}\ \emph {et~al.}(2016)\citenamefont {Chen},
  \citenamefont {Zhang}, \citenamefont {Wang},\ and\ \citenamefont
  {Zhao}}]{Chen_2016}%
  \BibitemOpen
  \bibfield  {author} {\bibinfo {author} {\bibfnamefont {S.}~\bibnamefont
  {Chen}}, \bibinfo {author} {\bibfnamefont {Y.}~\bibnamefont {Zhang}},
  \bibinfo {author} {\bibfnamefont {J.}~\bibnamefont {Wang}},\ and\ \bibinfo
  {author} {\bibfnamefont {H.}~\bibnamefont {Zhao}},\ }\bibfield  {title}
  {\bibinfo {title} {Key role of asymmetric interactions in low-dimensional
  heat transport},\ }\href {https://doi.org/10.1088/1742-5468/2016/03/033205}
  {\bibfield  {journal} {\bibinfo  {journal} {J. Stat. Mech: Theory Exp.}\
  }\textbf {\bibinfo {volume} {2016}},\ \bibinfo {pages} {033205} (\bibinfo
  {year} {2016})}\BibitemShut {NoStop}%
\bibitem [{\citenamefont {Jiang}\ and\ \citenamefont
  {Zhao}(2016)}]{Jiang_2016}%
  \BibitemOpen
  \bibfield  {author} {\bibinfo {author} {\bibfnamefont {J.}~\bibnamefont
  {Jiang}}\ and\ \bibinfo {author} {\bibfnamefont {H.}~\bibnamefont {Zhao}},\
  }\bibfield  {title} {\bibinfo {title} {Modulating thermal conduction by the
  axial strain},\ }\href {https://doi.org/10.1088/1742-5468/2016/09/093208}
  {\bibfield  {journal} {\bibinfo  {journal} {J. Stat. Mech: Theory Exp.}\
  }\textbf {\bibinfo {volume} {2016}},\ \bibinfo {pages} {093208} (\bibinfo
  {year} {2016})}\BibitemShut {NoStop}%
\bibitem [{\citenamefont {Chen}\ \emph
  {et~al.}(2014{\natexlab{b}})\citenamefont {Chen}, \citenamefont {Wang},
  \citenamefont {Casati},\ and\ \citenamefont {Benenti}}]{PhysRevE.90.032134}%
  \BibitemOpen
  \bibfield  {author} {\bibinfo {author} {\bibfnamefont {S.}~\bibnamefont
  {Chen}}, \bibinfo {author} {\bibfnamefont {J.}~\bibnamefont {Wang}}, \bibinfo
  {author} {\bibfnamefont {G.}~\bibnamefont {Casati}},\ and\ \bibinfo {author}
  {\bibfnamefont {G.}~\bibnamefont {Benenti}},\ }\bibfield  {title} {\bibinfo
  {title} {Nonintegrability and the {Fourier} heat conduction law},\ }\href
  {https://doi.org/10.1103/PhysRevE.90.032134} {\bibfield  {journal} {\bibinfo
  {journal} {Phys. Rev. E}\ }\textbf {\bibinfo {volume} {90}},\ \bibinfo
  {pages} {032134} (\bibinfo {year} {2014}{\natexlab{b}})}\BibitemShut
  {NoStop}%
\bibitem [{\citenamefont {Zhao}\ and\ \citenamefont
  {Wang}(2018)}]{PhysRevE.97.010103}%
  \BibitemOpen
  \bibfield  {author} {\bibinfo {author} {\bibfnamefont {H.}~\bibnamefont
  {Zhao}}\ and\ \bibinfo {author} {\bibfnamefont {W.-g.}\ \bibnamefont
  {Wang}},\ }\bibfield  {title} {\bibinfo {title} {Fourier heat conduction as a
  strong kinetic effect in one-dimensional hard-core gases},\ }\href
  {https://doi.org/10.1103/PhysRevE.97.010103} {\bibfield  {journal} {\bibinfo
  {journal} {Phys. Rev. E}\ }\textbf {\bibinfo {volume} {97}},\ \bibinfo
  {pages} {010103(R)} (\bibinfo {year} {2018})}\BibitemShut {NoStop}%
\bibitem [{\citenamefont {Lepri}\ \emph {et~al.}(2020)\citenamefont {Lepri},
  \citenamefont {Livi},\ and\ \citenamefont {Politi}}]{PhysRevLett.125.040604}%
  \BibitemOpen
  \bibfield  {author} {\bibinfo {author} {\bibfnamefont {S.}~\bibnamefont
  {Lepri}}, \bibinfo {author} {\bibfnamefont {R.}~\bibnamefont {Livi}},\ and\
  \bibinfo {author} {\bibfnamefont {A.}~\bibnamefont {Politi}},\ }\bibfield
  {title} {\bibinfo {title} {Too close to integrable: Crossover from normal to
  anomalous heat diffusion},\ }\href
  {https://doi.org/10.1103/PhysRevLett.125.040604} {\bibfield  {journal}
  {\bibinfo  {journal} {Phys. Rev. Lett.}\ }\textbf {\bibinfo {volume} {125}},\
  \bibinfo {pages} {040604} (\bibinfo {year} {2020})}\BibitemShut {NoStop}%
\bibitem [{\citenamefont {Casati}(1986)}]{casati1986energy}%
  \BibitemOpen
  \bibfield  {author} {\bibinfo {author} {\bibfnamefont {G.}~\bibnamefont
  {Casati}},\ }\bibfield  {title} {\bibinfo {title} {Energy transport and the
  {Fourier} heat law in classical systems},\ }\href
  {https://doi.org/10.1007/BF00735180} {\bibfield  {journal} {\bibinfo
  {journal} {Found. Phys.}\ }\textbf {\bibinfo {volume} {16}},\ \bibinfo
  {pages} {51} (\bibinfo {year} {1986})}\BibitemShut {NoStop}%
\bibitem [{\citenamefont {Casati}\ and\ \citenamefont
  {Ford}(1976)}]{casati1976computer}%
  \BibitemOpen
  \bibfield  {author} {\bibinfo {author} {\bibfnamefont {G.}~\bibnamefont
  {Casati}}\ and\ \bibinfo {author} {\bibfnamefont {J.}~\bibnamefont {Ford}},\
  }\bibfield  {title} {\bibinfo {title} {Computer study of ergodicity and
  mixing in a two-particle, hard point gas system},\ }\href
  {https://doi.org/https://doi.org/10.1016/0021-9991(76)90104-2} {\bibfield
  {journal} {\bibinfo  {journal} {J. Comput. Phys.}\ }\textbf {\bibinfo
  {volume} {20}},\ \bibinfo {pages} {97} (\bibinfo {year} {1976})}\BibitemShut
  {NoStop}%
\bibitem [{\citenamefont {Wang}\ \emph {et~al.}(2014)\citenamefont {Wang},
  \citenamefont {Casati},\ and\ \citenamefont {Prosen}}]{PhysRevE.89.042918}%
  \BibitemOpen
  \bibfield  {author} {\bibinfo {author} {\bibfnamefont {J.}~\bibnamefont
  {Wang}}, \bibinfo {author} {\bibfnamefont {G.}~\bibnamefont {Casati}},\ and\
  \bibinfo {author} {\bibfnamefont {T.}~\bibnamefont {Prosen}},\ }\bibfield
  {title} {\bibinfo {title} {Nonergodicity and localization of invariant
  measure for two colliding masses},\ }\href
  {https://doi.org/10.1103/PhysRevE.89.042918} {\bibfield  {journal} {\bibinfo
  {journal} {Phys. Rev. E}\ }\textbf {\bibinfo {volume} {89}},\ \bibinfo
  {pages} {042918} (\bibinfo {year} {2014})}\BibitemShut {NoStop}%
\bibitem [{\citenamefont {Boozer}(2011{\natexlab{a}})}]{Boozer_2011}%
  \BibitemOpen
  \bibfield  {author} {\bibinfo {author} {\bibfnamefont {A.~D.}\ \bibnamefont
  {Boozer}},\ }\bibfield  {title} {\bibinfo {title} {Boltzmann's {$H$}-theorem
  and the assumption of molecular chaos},\ }\href
  {https://doi.org/10.1088/0143-0807/32/5/027} {\bibfield  {journal} {\bibinfo
  {journal} {Eur. J. Phys.}\ }\textbf {\bibinfo {volume} {32}},\ \bibinfo
  {pages} {1391} (\bibinfo {year} {2011}{\natexlab{a}})}\BibitemShut {NoStop}%
\bibitem [{\citenamefont {Boozer}(2011{\natexlab{b}})}]{PhysRevE.84.031127}%
  \BibitemOpen
  \bibfield  {author} {\bibinfo {author} {\bibfnamefont {A.~D.}\ \bibnamefont
  {Boozer}},\ }\bibfield  {title} {\bibinfo {title} {Boltzmann equations for a
  binary one-dimensional ideal gas},\ }\href
  {https://doi.org/10.1103/PhysRevE.84.031127} {\bibfield  {journal} {\bibinfo
  {journal} {Phys. Rev. E}\ }\textbf {\bibinfo {volume} {84}},\ \bibinfo
  {pages} {031127} (\bibinfo {year} {2011}{\natexlab{b}})}\BibitemShut
  {NoStop}%
\bibitem [{\citenamefont {Dhar}(2001)}]{PhysRevLett.86.3554}%
  \BibitemOpen
  \bibfield  {author} {\bibinfo {author} {\bibfnamefont {A.}~\bibnamefont
  {Dhar}},\ }\bibfield  {title} {\bibinfo {title} {Heat conduction in a
  one-dimensional gas of elastically colliding particles of unequal masses},\
  }\href {https://doi.org/10.1103/PhysRevLett.86.3554} {\bibfield  {journal}
  {\bibinfo  {journal} {Phys. Rev. Lett.}\ }\textbf {\bibinfo {volume} {86}},\
  \bibinfo {pages} {3554} (\bibinfo {year} {2001})}\BibitemShut {NoStop}%
\bibitem [{\citenamefont {Garrido}\ \emph {et~al.}(2001)\citenamefont
  {Garrido}, \citenamefont {Hurtado},\ and\ \citenamefont
  {Nadrowski}}]{PhysRevLett.86.5486}%
  \BibitemOpen
  \bibfield  {author} {\bibinfo {author} {\bibfnamefont {P.~L.}\ \bibnamefont
  {Garrido}}, \bibinfo {author} {\bibfnamefont {P.~I.}\ \bibnamefont
  {Hurtado}},\ and\ \bibinfo {author} {\bibfnamefont {B.}~\bibnamefont
  {Nadrowski}},\ }\bibfield  {title} {\bibinfo {title} {Simple one-dimensional
  model of heat conduction which obeys {Fourier's} law},\ }\href
  {https://doi.org/10.1103/PhysRevLett.86.5486} {\bibfield  {journal} {\bibinfo
   {journal} {Phys. Rev. Lett.}\ }\textbf {\bibinfo {volume} {86}},\ \bibinfo
  {pages} {5486} (\bibinfo {year} {2001})}\BibitemShut {NoStop}%
\bibitem [{\citenamefont {Grassberger}\ \emph {et~al.}(2002)\citenamefont
  {Grassberger}, \citenamefont {Nadler},\ and\ \citenamefont
  {Yang}}]{PhysRevLett.89.180601}%
  \BibitemOpen
  \bibfield  {author} {\bibinfo {author} {\bibfnamefont {P.}~\bibnamefont
  {Grassberger}}, \bibinfo {author} {\bibfnamefont {W.}~\bibnamefont
  {Nadler}},\ and\ \bibinfo {author} {\bibfnamefont {L.}~\bibnamefont {Yang}},\
  }\bibfield  {title} {\bibinfo {title} {Heat conduction and entropy production
  in a one-dimensional hard-particle gas},\ }\href
  {https://doi.org/10.1103/PhysRevLett.89.180601} {\bibfield  {journal}
  {\bibinfo  {journal} {Phys. Rev. Lett.}\ }\textbf {\bibinfo {volume} {89}},\
  \bibinfo {pages} {180601} (\bibinfo {year} {2002})}\BibitemShut {NoStop}%
\bibitem [{\citenamefont {Casati}\ and\ \citenamefont
  {Prosen}(2003)}]{PhysRevE.67.015203}%
  \BibitemOpen
  \bibfield  {author} {\bibinfo {author} {\bibfnamefont {G.}~\bibnamefont
  {Casati}}\ and\ \bibinfo {author} {\bibfnamefont {T.}~\bibnamefont
  {Prosen}},\ }\bibfield  {title} {\bibinfo {title} {Anomalous heat conduction
  in a one-dimensional ideal gas},\ }\href
  {https://doi.org/10.1103/PhysRevE.67.015203} {\bibfield  {journal} {\bibinfo
  {journal} {Phys. Rev. E}\ }\textbf {\bibinfo {volume} {67}},\ \bibinfo
  {pages} {015203(R)} (\bibinfo {year} {2003})}\BibitemShut {NoStop}%
\bibitem [{\citenamefont {Cipriani}\ \emph {et~al.}(2005)\citenamefont
  {Cipriani}, \citenamefont {Denisov},\ and\ \citenamefont
  {Politi}}]{PhysRevLett.94.244301}%
  \BibitemOpen
  \bibfield  {author} {\bibinfo {author} {\bibfnamefont {P.}~\bibnamefont
  {Cipriani}}, \bibinfo {author} {\bibfnamefont {S.}~\bibnamefont {Denisov}},\
  and\ \bibinfo {author} {\bibfnamefont {A.}~\bibnamefont {Politi}},\
  }\bibfield  {title} {\bibinfo {title} {From anomalous energy diffusion to
  {Levy} walks and heat conductivity in one-dimensional systems},\ }\href
  {https://doi.org/10.1103/PhysRevLett.94.244301} {\bibfield  {journal}
  {\bibinfo  {journal} {Phys. Rev. Lett.}\ }\textbf {\bibinfo {volume} {94}},\
  \bibinfo {pages} {244301} (\bibinfo {year} {2005})}\BibitemShut {NoStop}%
\bibitem [{\citenamefont {Hurtado}\ and\ \citenamefont
  {Garrido}(2020)}]{hurtado2020simulations}%
  \BibitemOpen
  \bibfield  {author} {\bibinfo {author} {\bibfnamefont {P.~I.}\ \bibnamefont
  {Hurtado}}\ and\ \bibinfo {author} {\bibfnamefont {P.~L.}\ \bibnamefont
  {Garrido}},\ }\bibfield  {title} {\bibinfo {title} {Simulations of transport
  in hard particle systems},\ }\href
  {https://doi.org/10.1007/s10955-019-02469-z} {\bibfield  {journal} {\bibinfo
  {journal} {J. Stat. Phys.}\ }\textbf {\bibinfo {volume} {180}},\ \bibinfo
  {pages} {474} (\bibinfo {year} {2020})}\BibitemShut {NoStop}%
\bibitem [{\citenamefont {Mendl}\ and\ \citenamefont
  {Spohn}(2014)}]{PhysRevE.90.012147}%
  \BibitemOpen
  \bibfield  {author} {\bibinfo {author} {\bibfnamefont {C.~B.}\ \bibnamefont
  {Mendl}}\ and\ \bibinfo {author} {\bibfnamefont {H.}~\bibnamefont {Spohn}},\
  }\bibfield  {title} {\bibinfo {title} {Equilibrium time-correlation functions
  for one-dimensional hard-point systems},\ }\href
  {https://doi.org/10.1103/PhysRevE.90.012147} {\bibfield  {journal} {\bibinfo
  {journal} {Phys. Rev. E}\ }\textbf {\bibinfo {volume} {90}},\ \bibinfo
  {pages} {012147} (\bibinfo {year} {2014})}\BibitemShut {NoStop}%
\bibitem [{\citenamefont {Chakraborti}\ \emph {et~al.}(2021)\citenamefont
  {Chakraborti}, \citenamefont {Ganapa}, \citenamefont {Krapivsky},\ and\
  \citenamefont {Dhar}}]{PhysRevLett.126.244503}%
  \BibitemOpen
  \bibfield  {author} {\bibinfo {author} {\bibfnamefont {S.}~\bibnamefont
  {Chakraborti}}, \bibinfo {author} {\bibfnamefont {S.}~\bibnamefont {Ganapa}},
  \bibinfo {author} {\bibfnamefont {P.~L.}\ \bibnamefont {Krapivsky}},\ and\
  \bibinfo {author} {\bibfnamefont {A.}~\bibnamefont {Dhar}},\ }\bibfield
  {title} {\bibinfo {title} {Blast in a one-dimensional cold gas: From
  {Newtonian} dynamics to hydrodynamics},\ }\href
  {https://doi.org/10.1103/PhysRevLett.126.244503} {\bibfield  {journal}
  {\bibinfo  {journal} {Phys. Rev. Lett.}\ }\textbf {\bibinfo {volume} {126}},\
  \bibinfo {pages} {244503} (\bibinfo {year} {2021})}\BibitemShut {NoStop}%
\bibitem [{\citenamefont {Bertini}\ \emph {et~al.}(2021)\citenamefont
  {Bertini}, \citenamefont {Heidrich-Meisner}, \citenamefont {Karrasch},
  \citenamefont {Prosen}, \citenamefont {Steinigeweg},\ and\ \citenamefont
  {\ifmmode \check{Z}\else \v{Z}\fi{}nidari\ifmmode~\check{c}\else
  \v{c}\fi{}}}]{RevModPhys.93.025003}%
  \BibitemOpen
  \bibfield  {author} {\bibinfo {author} {\bibfnamefont {B.}~\bibnamefont
  {Bertini}}, \bibinfo {author} {\bibfnamefont {F.}~\bibnamefont
  {Heidrich-Meisner}}, \bibinfo {author} {\bibfnamefont {C.}~\bibnamefont
  {Karrasch}}, \bibinfo {author} {\bibfnamefont {T.}~\bibnamefont {Prosen}},
  \bibinfo {author} {\bibfnamefont {R.}~\bibnamefont {Steinigeweg}},\ and\
  \bibinfo {author} {\bibfnamefont {M.}~\bibnamefont {\ifmmode \check{Z}\else
  \v{Z}\fi{}nidari\ifmmode~\check{c}\else \v{c}\fi{}}},\ }\bibfield  {title}
  {\bibinfo {title} {Finite-temperature transport in one-dimensional quantum
  lattice models},\ }\href {https://doi.org/10.1103/RevModPhys.93.025003}
  {\bibfield  {journal} {\bibinfo  {journal} {Rev. Mod. Phys.}\ }\textbf
  {\bibinfo {volume} {93}},\ \bibinfo {pages} {025003} (\bibinfo {year}
  {2021})}\BibitemShut {NoStop}%
\bibitem [{\citenamefont {Zuckerwar}(2002)}]{wong2002handbook}%
  \BibitemOpen
  \bibfield  {author} {\bibinfo {author} {\bibfnamefont {A.~J.}\ \bibnamefont
  {Zuckerwar}},\ }\href@noop {} {\emph {\bibinfo {title} {Handbook of the Speed
  of Sound in Real Gases}}}\ (\bibinfo  {publisher} {Academic Press},\ \bibinfo
  {year} {2002})\BibitemShut {NoStop}%
\bibitem [{tim()}]{timeTemp}%
  \BibitemOpen
  \href@noop {} {}\bibinfo {note} {Consider that time $t\sim L/v$ and the
  kinctic temperature $T=\langle mv^2\rangle$, so that $t\sim L/\sqrt{T/m}\sim
  T^{-1/2}$.}\BibitemShut {Stop}%
\bibitem [{Fud()}]{Fudmass2}%
  \BibitemOpen
  \href@noop {} {}\bibinfo {note} {{W. Fu, Z. Wang, Y. Wang, Y. Zhang, and H.
  Zhao} (to be published)}\BibitemShut {NoStop}%
\bibitem [{\citenamefont {Bogoliubov}(1962)}]{bogoliubov1962problems}%
  \BibitemOpen
  \bibfield  {author} {\bibinfo {author} {\bibfnamefont {N.~N.}\ \bibnamefont
  {Bogoliubov}},\ }\href@noop {} {\emph {\bibinfo {title} {Problems of Dynamic
  Theory in Statistical Physics}}}\ (\bibinfo  {publisher} {Studies in
  Statistical Mechanics, Vol. I, North-Holland, Amsterdam},\ \bibinfo {year}
  {1962})\BibitemShut {NoStop}%
\bibitem [{\citenamefont {Liboff}(1985)}]{PhysRevA.31.1883}%
  \BibitemOpen
  \bibfield  {author} {\bibinfo {author} {\bibfnamefont {R.~L.}\ \bibnamefont
  {Liboff}},\ }\bibfield  {title} {\bibinfo {title} {Generalized {Bogoliubov}
  hypothesis for dense fluids},\ }\href
  {https://doi.org/10.1103/PhysRevA.31.1883} {\bibfield  {journal} {\bibinfo
  {journal} {Phys. Rev. A}\ }\textbf {\bibinfo {volume} {31}},\ \bibinfo
  {pages} {1883} (\bibinfo {year} {1985})}\BibitemShut {NoStop}%
\bibitem [{\citenamefont {Mitropolskii}\ and\ \citenamefont
  {Petrina}(1993)}]{mitropolskii1993nn}%
  \BibitemOpen
  \bibfield  {author} {\bibinfo {author} {\bibfnamefont {Y.~A.}\ \bibnamefont
  {Mitropolskii}}\ and\ \bibinfo {author} {\bibfnamefont {D.~Y.}\ \bibnamefont
  {Petrina}},\ }\bibfield  {title} {\bibinfo {title} {On {N. N. Bogolyubov's}
  works in classical and quantum statistical mechanics},\ }\href
  {https://doi.org/10.1007/BF01060975} {\bibfield  {journal} {\bibinfo
  {journal} {Ukr. Math. J.}\ }\textbf {\bibinfo {volume} {45}},\ \bibinfo
  {pages} {171} (\bibinfo {year} {1993})}\BibitemShut {NoStop}%
\bibitem [{\citenamefont {(Jr.)}\ and\ \citenamefont
  {Sankovich}(1994)}]{Bogolyubov_Jr_1994}%
  \BibitemOpen
  \bibfield  {author} {\bibinfo {author} {\bibfnamefont {N.~N.~B.}\
  \bibnamefont {(Jr.)}}\ and\ \bibinfo {author} {\bibfnamefont {D.~P.}\
  \bibnamefont {Sankovich}},\ }\bibfield  {title} {\bibinfo {title} {{N.
  N. Bogolyubov} and statistical mechanics},\ }\href
  {https://doi.org/10.1070/RM1994v049n05ABEH002419} {\bibfield  {journal}
  {\bibinfo  {journal} {Russ. Math. Surv.}\ }\textbf {\bibinfo {volume} {49}},\
  \bibinfo {pages} {19} (\bibinfo {year} {1994})}\BibitemShut {NoStop}%
\bibitem [{\citenamefont {Boltzmann}(1964)}]{boltzmann1964lectures}%
  \BibitemOpen
  \bibfield  {author} {\bibinfo {author} {\bibfnamefont {L.}~\bibnamefont
  {Boltzmann}},\ }\href@noop {} {\emph {\bibinfo {title} {Lectures on gas
  theory}}}\ (\bibinfo  {publisher} {University of California Press},\ \bibinfo
  {year} {1964})\BibitemShut {NoStop}%
\bibitem [{\citenamefont {Chapman}\ and\ \citenamefont
  {Cowling}(1990)}]{chapman1990mathematical}%
  \BibitemOpen
  \bibfield  {author} {\bibinfo {author} {\bibfnamefont {S.}~\bibnamefont
  {Chapman}}\ and\ \bibinfo {author} {\bibfnamefont {T.~G.}\ \bibnamefont
  {Cowling}},\ }\href@noop {} {\emph {\bibinfo {title} {The mathematical theory
  of non-uniform gases: an account of the kinetic theory of viscosity, thermal
  conduction and diffusion in gases}}},\ \bibinfo {edition} {3rd}\ ed.\
  (\bibinfo  {publisher} {Cambridge university press},\ \bibinfo {year}
  {1990})\BibitemShut {NoStop}%
\bibitem [{\citenamefont {Ehrenfest}\ and\ \citenamefont
  {Ehrenfest}(1990)}]{ehrenfest1990conceptual}%
  \BibitemOpen
  \bibfield  {author} {\bibinfo {author} {\bibfnamefont {P.}~\bibnamefont
  {Ehrenfest}}\ and\ \bibinfo {author} {\bibfnamefont {T.}~\bibnamefont
  {Ehrenfest}},\ }\href@noop {} {\emph {\bibinfo {title} {The Conceptual
  Foundations of the Statistical Approach in Mechanics}}}\ (\bibinfo
  {publisher} {Dover Publications, Inc},\ \bibinfo {address} {New York},\
  \bibinfo {year} {1990})\BibitemShut {NoStop}%
\bibitem [{\citenamefont {Brown}\ \emph {et~al.}(2009)\citenamefont {Brown},
  \citenamefont {Myrvold},\ and\ \citenamefont {Uffink}}]{Brown2009}%
  \BibitemOpen
  \bibfield  {author} {\bibinfo {author} {\bibfnamefont {H.~R.}\ \bibnamefont
  {Brown}}, \bibinfo {author} {\bibfnamefont {W.}~\bibnamefont {Myrvold}},\
  and\ \bibinfo {author} {\bibfnamefont {J.}~\bibnamefont {Uffink}},\
  }\bibfield  {title} {\bibinfo {title} {Boltzmann's {$H$}-theorem, its
  discontents, and the birth of statistical mechanics},\ }\href
  {https://doi.org/https://doi.org/10.1016/j.shpsb.2009.03.003} {\bibfield
  {journal} {\bibinfo  {journal} {Stud. Hist. Philos. M. P.}\ }\textbf
  {\bibinfo {volume} {40}},\ \bibinfo {pages} {174} (\bibinfo {year}
  {2009})}\BibitemShut {NoStop}%
\bibitem [{\citenamefont {Wegner}(1980)}]{Wegner1980}%
  \BibitemOpen
  \bibfield  {author} {\bibinfo {author} {\bibfnamefont {F.}~\bibnamefont
  {Wegner}},\ }\bibfield  {title} {\bibinfo {title} {Inverse participation
  ratio in $2+\epsilon$ dimensions},\ }\href
  {https://doi.org/10.1007/BF01325284} {\bibfield  {journal} {\bibinfo
  {journal} {Z. Phys. B Condens. Matter}\ }\textbf {\bibinfo {volume} {36}},\
  \bibinfo {pages} {209} (\bibinfo {year} {1980})}\BibitemShut {NoStop}%
\bibitem [{IPR()}]{IPR_Var}%
  \BibitemOpen
  \href@noop {} {}\bibinfo {note} {The fluctuations of local energy can be
  defined as $\Delta
  E(t)=\sum_{i=1}^{N}\left(\mathcal{E}_i(t)-E/N\right)^2=\sum_{i=1}^{N}\mathcal{E}_i(t)^2-E^2/N$,
  where $E=\sum_{i=1}^N\mathcal{E}_i$ is the total energy.}\BibitemShut {Stop}%
\bibitem [{\citenamefont {Grad}(1949)}]{grad1949kinetic}%
  \BibitemOpen
  \bibfield  {author} {\bibinfo {author} {\bibfnamefont {H.}~\bibnamefont
  {Grad}},\ }\bibfield  {title} {\bibinfo {title} {On the kinetic theory of
  rarefied gases},\ }\href {https://doi.org/10.1002/cpa.3160020403} {\bibfield
  {journal} {\bibinfo  {journal} {Comm. Pure Appl. Math.}\ }\textbf {\bibinfo
  {volume} {2}},\ \bibinfo {pages} {331} (\bibinfo {year} {1949})}\BibitemShut
  {NoStop}%
\bibitem [{\citenamefont {Qian}\ \emph {et~al.}(2016)\citenamefont {Qian},
  \citenamefont {Ao}, \citenamefont {Tu},\ and\ \citenamefont
  {Wang}}]{QIAN2016153}%
  \BibitemOpen
  \bibfield  {author} {\bibinfo {author} {\bibfnamefont {H.}~\bibnamefont
  {Qian}}, \bibinfo {author} {\bibfnamefont {P.}~\bibnamefont {Ao}}, \bibinfo
  {author} {\bibfnamefont {Y.}~\bibnamefont {Tu}},\ and\ \bibinfo {author}
  {\bibfnamefont {J.}~\bibnamefont {Wang}},\ }\bibfield  {title} {\bibinfo
  {title} {A framework towards understanding mesoscopic phenomena: Emergent
  unpredictability, symmetry breaking and dynamics across scales},\ }\href
  {https://doi.org/https://doi.org/10.1016/j.cplett.2016.10.059} {\bibfield
  {journal} {\bibinfo  {journal} {Chem. Phys. Lett.}\ }\textbf {\bibinfo
  {volume} {665}},\ \bibinfo {pages} {153} (\bibinfo {year}
  {2016})}\BibitemShut {NoStop}%
\bibitem [{\citenamefont {Onsager}(1931)}]{PhysRev.37.405}%
  \BibitemOpen
  \bibfield  {author} {\bibinfo {author} {\bibfnamefont {L.}~\bibnamefont
  {Onsager}},\ }\bibfield  {title} {\bibinfo {title} {Reciprocal relations in
  irreversible processes. {I}.},\ }\href
  {https://doi.org/10.1103/PhysRev.37.405} {\bibfield  {journal} {\bibinfo
  {journal} {Phys. Rev.}\ }\textbf {\bibinfo {volume} {37}},\ \bibinfo {pages}
  {405} (\bibinfo {year} {1931})}\BibitemShut {NoStop}%
\bibitem [{\citenamefont {Kubo}(1957)}]{kubo1957statistical}%
  \BibitemOpen
  \bibfield  {author} {\bibinfo {author} {\bibfnamefont {R.}~\bibnamefont
  {Kubo}},\ }\bibfield  {title} {\bibinfo {title} {Statistical-mechanical
  theory of irreversible processes. i. general theory and simple applications
  to magnetic and conduction problems},\ }\href
  {https://doi.org/10.1143/JPSJ.12.570} {\bibfield  {journal} {\bibinfo
  {journal} {J. Phys. Soc. Jpn.}\ }\textbf {\bibinfo {volume} {12}},\ \bibinfo
  {pages} {570} (\bibinfo {year} {1957})}\BibitemShut {NoStop}%
\bibitem [{\citenamefont {Kubo}\ \emph {et~al.}(1957)\citenamefont {Kubo},
  \citenamefont {Yokota},\ and\ \citenamefont {Nakajima}}]{kubo1957stat2}%
  \BibitemOpen
  \bibfield  {author} {\bibinfo {author} {\bibfnamefont {R.}~\bibnamefont
  {Kubo}}, \bibinfo {author} {\bibfnamefont {M.}~\bibnamefont {Yokota}},\ and\
  \bibinfo {author} {\bibfnamefont {S.}~\bibnamefont {Nakajima}},\ }\bibfield
  {title} {\bibinfo {title} {Statistical-mechanical theory of irreversible
  processes. ii. response to thermal disturbance},\ }\href
  {https://doi.org/10.1143/JPSJ.12.1203} {\bibfield  {journal} {\bibinfo
  {journal} {J. Phys. Soc. Jpn.}\ }\textbf {\bibinfo {volume} {12}},\ \bibinfo
  {pages} {1203} (\bibinfo {year} {1957})}\BibitemShut {NoStop}%
\bibitem [{\citenamefont {Landau}\ and\ \citenamefont
  {Lifshitz}(1980)}]{landau2013statistical}%
  \BibitemOpen
  \bibfield  {author} {\bibinfo {author} {\bibfnamefont {L.~D.}\ \bibnamefont
  {Landau}}\ and\ \bibinfo {author} {\bibfnamefont {E.~M.}\ \bibnamefont
  {Lifshitz}},\ }\href@noop {} {\emph {\bibinfo {title} {Statistical
  Physics}}},\ \bibinfo {edition} {3rd}\ ed.\ (\bibinfo  {publisher} {Pergamon
  Press},\ \bibinfo {address} {Oxford},\ \bibinfo {year} {1980})\BibitemShut
  {NoStop}%
\bibitem [{kap()}]{kappaForDzeros}%
  \BibitemOpen
  \href@noop {} {}\bibinfo {note} {According to the kinetic theory of gases,
  $\kappa=c v_{\rm s} l/d$, where $c$ represents the heat capacity, $v_{\rm s}$
  denotes the sound velocity, $l$ stands for the mean free path, and $d$
  indicates the dimension of the system. For 1D diatomic hard-point gases
  studied in this context, we have $c=1/2$, $v_{\rm s}=\sqrt{3T}$, and $d=1$.
  It should be noted that here we interpret $l=v_{\rm s}\tau$ as the mean-free
  path of energy wave peaks which propagate ballistically at the speed of
  sound. In this system, particles serve as exclusive energy carriers and
  possess their own mean-free path equivalent to mean particle spacing (here
  being unit). However, when particles with equal mass collide, they directly
  exchange energy without scattering or disrupting the propagation of energy
  wave packets. This behavior is akin to particles colliding through each other
  in a classical sense. The distinguishability of classical particles allows us
  to differentiate between them. If particles were indistinguishable, their
  collision would resemble quantum tunneling. Here $\tau$ represents mean-free
  path time and solely depends on system size. Consequently,
  $\kappa(\tau)=\frac{3T}{2}\tau$. Interestingly enough, the coefficient
  preceding $\tau$ in expression~(\ref{eq-kappa-our0}) is exactly identical to
  that obtained here for $\delta=0$.}\BibitemShut {Stop}%
\bibitem [{\citenamefont {Wang}\ \emph {et~al.}(2013)\citenamefont {Wang},
  \citenamefont {Hu},\ and\ \citenamefont {Li}}]{PhysRevE.88.052112}%
  \BibitemOpen
  \bibfield  {author} {\bibinfo {author} {\bibfnamefont {L.}~\bibnamefont
  {Wang}}, \bibinfo {author} {\bibfnamefont {B.}~\bibnamefont {Hu}},\ and\
  \bibinfo {author} {\bibfnamefont {B.}~\bibnamefont {Li}},\ }\bibfield
  {title} {\bibinfo {title} {Validity of fourier's law in one-dimensional
  momentum-conserving lattices with asymmetric interparticle interactions},\
  }\href {https://doi.org/10.1103/PhysRevE.88.052112} {\bibfield  {journal}
  {\bibinfo  {journal} {Phys. Rev. E}\ }\textbf {\bibinfo {volume} {88}},\
  \bibinfo {pages} {052112} (\bibinfo {year} {2013})}\BibitemShut {NoStop}%
\bibitem [{\citenamefont {Das}\ \emph {et~al.}(2014)\citenamefont {Das},
  \citenamefont {Dhar},\ and\ \citenamefont {Narayan}}]{Das2014}%
  \BibitemOpen
  \bibfield  {author} {\bibinfo {author} {\bibfnamefont {S.~G.}\ \bibnamefont
  {Das}}, \bibinfo {author} {\bibfnamefont {A.}~\bibnamefont {Dhar}},\ and\
  \bibinfo {author} {\bibfnamefont {O.}~\bibnamefont {Narayan}},\ }\bibfield
  {title} {\bibinfo {title} {Heat conduction in the $\alpha$-$\beta$
  fermi-pasta-ulam chain},\ }\href {https://doi.org/10.1007/s10955-013-0871-0}
  {\bibfield  {journal} {\bibinfo  {journal} {Journal of Statistical Physics}\
  }\textbf {\bibinfo {volume} {154}},\ \bibinfo {pages} {204} (\bibinfo {year}
  {2014})}\BibitemShut {NoStop}%
\bibitem [{\citenamefont {Ghosh}\ \emph {et~al.}(2022)\citenamefont {Ghosh},
  \citenamefont {Kusiak},\ and\ \citenamefont {Battaglia}}]{Ghosh_2022}%
  \BibitemOpen
  \bibfield  {author} {\bibinfo {author} {\bibfnamefont {K.}~\bibnamefont
  {Ghosh}}, \bibinfo {author} {\bibfnamefont {A.}~\bibnamefont {Kusiak}},\ and\
  \bibinfo {author} {\bibfnamefont {J.~L.}\ \bibnamefont {Battaglia}},\
  }\bibfield  {title} {\bibinfo {title} {Phonon hydrodynamics in crystalline
  materials},\ }\href {https://doi.org/10.1088/1361-648X/ac718a} {\bibfield
  {journal} {\bibinfo  {journal} {J. Phys. Condens. Matter}\ }\textbf {\bibinfo
  {volume} {34}},\ \bibinfo {pages} {323001} (\bibinfo {year}
  {2022})}\BibitemShut {NoStop}%
\bibitem [{\citenamefont {Gershgorin}\ \emph {et~al.}(2005)\citenamefont
  {Gershgorin}, \citenamefont {Lvov},\ and\ \citenamefont
  {Cai}}]{PhysRevLett.95.264302}%
  \BibitemOpen
  \bibfield  {author} {\bibinfo {author} {\bibfnamefont {B.}~\bibnamefont
  {Gershgorin}}, \bibinfo {author} {\bibfnamefont {Y.~V.}\ \bibnamefont
  {Lvov}},\ and\ \bibinfo {author} {\bibfnamefont {D.}~\bibnamefont {Cai}},\
  }\bibfield  {title} {\bibinfo {title} {Renormalized waves and discrete
  breathers in $\ensuremath{\beta}$-fermi-pasta-ulam chains},\ }\href
  {https://doi.org/10.1103/PhysRevLett.95.264302} {\bibfield  {journal}
  {\bibinfo  {journal} {Phys. Rev. Lett.}\ }\textbf {\bibinfo {volume} {95}},\
  \bibinfo {pages} {264302} (\bibinfo {year} {2005})}\BibitemShut {NoStop}%
\bibitem [{\citenamefont {Onorato}\ \emph {et~al.}(2015)\citenamefont
  {Onorato}, \citenamefont {Vozella}, \citenamefont {Proment},\ and\
  \citenamefont {Lvov}}]{Onorato4208}%
  \BibitemOpen
  \bibfield  {author} {\bibinfo {author} {\bibfnamefont {M.}~\bibnamefont
  {Onorato}}, \bibinfo {author} {\bibfnamefont {L.}~\bibnamefont {Vozella}},
  \bibinfo {author} {\bibfnamefont {D.}~\bibnamefont {Proment}},\ and\ \bibinfo
  {author} {\bibfnamefont {Y.~V.}\ \bibnamefont {Lvov}},\ }\bibfield  {title}
  {\bibinfo {title} {Route to thermalization in the $\alpha$-{Fermi-Pasta-Ulam}
  system},\ }\href {https://doi.org/10.1073/pnas.1404397112} {\bibfield
  {journal} {\bibinfo  {journal} {Proc. Natl. Acad. Sci. U.S.A.}\ }\textbf
  {\bibinfo {volume} {112}},\ \bibinfo {pages} {4208} (\bibinfo {year}
  {2015})}\BibitemShut {NoStop}%
\bibitem [{\citenamefont {Lvov}\ and\ \citenamefont
  {Onorato}(2018)}]{PhysRevLett.120.144301}%
  \BibitemOpen
  \bibfield  {author} {\bibinfo {author} {\bibfnamefont {Y.~V.}\ \bibnamefont
  {Lvov}}\ and\ \bibinfo {author} {\bibfnamefont {M.}~\bibnamefont {Onorato}},\
  }\bibfield  {title} {\bibinfo {title} {Double scaling in the relaxation time
  in the $\beta$-{Fermi-Pasta-Ulam-Tsingou} model},\ }\href
  {https://doi.org/10.1103/PhysRevLett.120.144301} {\bibfield  {journal}
  {\bibinfo  {journal} {Phys. Rev. Lett.}\ }\textbf {\bibinfo {volume} {120}},\
  \bibinfo {pages} {144301} (\bibinfo {year} {2018})}\BibitemShut {NoStop}%
\bibitem [{\citenamefont {Helfand}(1960)}]{PhysRev.119.1}%
  \BibitemOpen
  \bibfield  {author} {\bibinfo {author} {\bibfnamefont {E.}~\bibnamefont
  {Helfand}},\ }\bibfield  {title} {\bibinfo {title} {Transport coefficients
  from dissipation in a canonical ensemble},\ }\href
  {https://doi.org/10.1103/PhysRev.119.1} {\bibfield  {journal} {\bibinfo
  {journal} {Phys. Rev.}\ }\textbf {\bibinfo {volume} {119}},\ \bibinfo {pages}
  {1} (\bibinfo {year} {1960})}\BibitemShut {NoStop}%
\bibitem [{\citenamefont {Benenti}\ \emph {et~al.}(2013)\citenamefont
  {Benenti}, \citenamefont {Casati},\ and\ \citenamefont
  {Wang}}]{PhysRevLett.110.070604}%
  \BibitemOpen
  \bibfield  {author} {\bibinfo {author} {\bibfnamefont {G.}~\bibnamefont
  {Benenti}}, \bibinfo {author} {\bibfnamefont {G.}~\bibnamefont {Casati}},\
  and\ \bibinfo {author} {\bibfnamefont {J.}~\bibnamefont {Wang}},\ }\bibfield
  {title} {\bibinfo {title} {Conservation laws and thermodynamic
  efficiencies},\ }\href {https://doi.org/10.1103/PhysRevLett.110.070604}
  {\bibfield  {journal} {\bibinfo  {journal} {Phys. Rev. Lett.}\ }\textbf
  {\bibinfo {volume} {110}},\ \bibinfo {pages} {070604} (\bibinfo {year}
  {2013})}\BibitemShut {NoStop}%
\bibitem [{\citenamefont {Chen}\ \emph {et~al.}(2015)\citenamefont {Chen},
  \citenamefont {Wang}, \citenamefont {Casati},\ and\ \citenamefont
  {Benenti}}]{Chen2015}%
  \BibitemOpen
  \bibfield  {author} {\bibinfo {author} {\bibfnamefont {S.}~\bibnamefont
  {Chen}}, \bibinfo {author} {\bibfnamefont {J.}~\bibnamefont {Wang}}, \bibinfo
  {author} {\bibfnamefont {G.}~\bibnamefont {Casati}},\ and\ \bibinfo {author}
  {\bibfnamefont {G.}~\bibnamefont {Benenti}},\ }\bibfield  {title} {\bibinfo
  {title} {Thermoelectricity of interacting particles: A numerical approach},\
  }\href {https://doi.org/10.1103/PhysRevE.92.032139} {\bibfield  {journal}
  {\bibinfo  {journal} {Phys. Rev. E}\ }\textbf {\bibinfo {volume} {92}},\
  \bibinfo {pages} {032139} (\bibinfo {year} {2015})}\BibitemShut {NoStop}%
\bibitem [{\citenamefont {Luo}\ \emph {et~al.}(2018)\citenamefont {Luo},
  \citenamefont {Benenti}, \citenamefont {Casati},\ and\ \citenamefont
  {Wang}}]{PhysRevLett.121.080602}%
  \BibitemOpen
  \bibfield  {author} {\bibinfo {author} {\bibfnamefont {R.}~\bibnamefont
  {Luo}}, \bibinfo {author} {\bibfnamefont {G.}~\bibnamefont {Benenti}},
  \bibinfo {author} {\bibfnamefont {G.}~\bibnamefont {Casati}},\ and\ \bibinfo
  {author} {\bibfnamefont {J.}~\bibnamefont {Wang}},\ }\bibfield  {title}
  {\bibinfo {title} {Thermodynamic bound on heat-to-power conversion},\ }\href
  {https://doi.org/10.1103/PhysRevLett.121.080602} {\bibfield  {journal}
  {\bibinfo  {journal} {Phys. Rev. Lett.}\ }\textbf {\bibinfo {volume} {121}},\
  \bibinfo {pages} {080602} (\bibinfo {year} {2018})}\BibitemShut {NoStop}%
\end{thebibliography}%

\end{document}